\documentclass[12pt,a4paper,dvips]{article}
\usepackage{a4p}
\usepackage{cite,mcite}
\usepackage{graphicx}
\usepackage{physics}
\usepackage{l3_title,ifthen,Lep}
\journalname{Phys. Lett. B}
\date{March 26, 2003}
\preprint{2003-013}
\newlength{\capindent}
\newlength{\capwidth}
\newlength{\figwidth}
\newcommand{\icaption}[2][!*!,!]{\hspace*{\capindent}%
  \begin{minipage}{\capwidth}
    \ifthenelse{\equal{#1}{!*!,!}}%
      {\caption{#2}}%
      {\caption[#1]{#2}}
  \end{minipage}}
\newcommand{\notE}{\rm\hspace{-0.7ex}\not\hspace{-0.2ex} E}
\begin{document}
\begin{titlepage}
\title{Search for Excited Leptons at LEP }

\author{The L3 Collaboration}

%
%
\begin{abstract}

A search for charged and neutral excited leptons is performed in 217
pb$^{-1}$ of $\epem$ collision data collected with the L3 detector at
LEP at centre-of-mass energies from $202\GeV$ up to $209\GeV$. The
pair- and single-production mechanisms of excited electrons, muons and
taus, as well as of excited electron-, muon- and tau-neutrinos, are
investigated and no signals are detected. Combining with L3 results
from searches at lower centre-of-mass energies, gives improved limits
on the masses and couplings of excited leptons.
\end{abstract}

\submitted
\end{titlepage}

%
%
\section{Introduction} 

Charged ($\rm e^*$, $\mu^*$, and $\tau^*$) and neutral ($\nu_{\rm
e}^*$, $\nu_{\mu}^*$ and $\nu_{\tau}^*$) excited leptons are predicted
by composite models where leptons and quarks have
substructure\cite{hagi,neus,yellow}. These models address fundamental
questions left open by the Standard Model\cite{SM}, such as the number
of families and the fermion mass values.

High energy electron-positron annihilations constitute an excellent
environment for the search for excited leptons and several searches
have been carried out at LEP~\cite{L3_189,L3_192,LEP}.  Searches for
excited electrons and neutrinos were also performed at the
HERA\cite{HERA} $\rm ep$ collider.  This Letter describes the
extension of previous L3 searches~\cite{L3_189,L3_192} to the highest
centre-of-mass energies, $\sqrt{s}$, attained by the LEP machine in its last year
of operation, $\sqrt{s}=202-209\GeV$.

Excited leptons are studied within a model~\cite{hagi} 
in which they are 
described as isospin doublets with left and right handed components: 
\begin{equation}
  {\rm L}^* =
  \left(\begin{array}{c} \nu^* \\ \ell^* \end{array}\right)_L + 
  \left(\begin{array}{c} \nu^* \\ \ell^* \end{array}\right)_R,
\end{equation}
with $\rm \ell = e, \mu, \tau$ and $\rm \nu = \nu_{e}, \nu_{\mu}, \nu_{\tau}$.
These excited leptons are accessible at LEP through pair production, $\rm e^+e^-
\rightarrow \ell^*\ell^*$, $\nu^*\nu^*$, or single production, $ \rm e^+e^-
\rightarrow \ell\ell^*$, $\nu\nu^*$.
Pair-production searches are sensitive to excited leptons of mass up
to values close to the kinematic limit $\sqrt{s}/2$, while the
single-production mechanism extends the search potential up to masses
close to $\sqrt{s}$.

The pair production of excited leptons is described by the effective
Lagrangian: 

\begin{equation}
{\cal L}_{{\rm L}^*{\rm L}^*} = \bar{\rm L}^* \gamma^{\mu} \left( g  
{\vec\tau \over 2} \vec W_{\mu} + {g'} Y B_{\mu} \right) {\rm L}^*. 
\label{eq:2}
\end{equation}
while the single production of excited leptons and their decay into
Standard Model leptons is modeled by:

\begin{equation}
{\cal L}_{{\rm L}^*{\rm L}} = 
{\bar {\rm L}^*} \sigma^{\mu\nu}  
\left( g { f  \over \Lambda}  \frac{\vec\tau}{2} \partial_{\mu}{\vec W_{\nu}} + 
       g' {f'  \over \Lambda} Y \partial_{\mu}B_{\nu} \right)  
       \frac{1-\gamma^5}{2} {\rm L} + ({\rm hermitian\,\,\, conjugate}).
\label{eq:3}
\end{equation}
In these expressions, $\gamma^{\mu}$ are the Dirac matrices,
$\sigma^{\mu\nu} =  i
[\gamma^{\mu},\gamma^{\nu}]/2$, $g$ and $g'$ are the Standard Model
SU(2) and U(1) coupling constants, $\vec\tau$ denotes the Pauli
matrices, $Y=-1/2$ is the hypercharge and $\vec W$ and $B$ are the
gauge fields associated with the SU(2) and U(1) groups respectively. L
denotes the Standard Model leptons, $\Lambda$ is the scale of the New
Physics responsible for the existence of excited leptons and $f$ and
$f'$ scale the SU(2) and U(1) couplings, respectively.

The cross section for pair production of excited leptons depends only
on their mass and on $\sqrt{s}$. As an example, for masses of the
excited leptons of $101\GeV$, the cross sections at $\sqrt{s}=206\GeV$
for $\rm e^+e^- \rightarrow \mu^*\mu^*$ and $\rm e^+e^- \rightarrow
\nu^*_\mu\nu^*_\mu$ are 0.6\,pb and 0.3\,pb respectively.  The
single-production cross section for excited leptons depends also
(equation~\ref{eq:3}) on $f/\Lambda$ and $f'/\Lambda$.

Equation~\ref{eq:3} also describes the decay of excited leptons into
Standard Model leptons in association with a photon or a gauge boson.
Three decays are possible: radiative decays, $\rm \ell^* \rightarrow
\ell \gamma$ and $\rm \nu^* \rightarrow \nu \gamma$, charged-current
decays, $\rm \ell^* \rightarrow \nu W$ and $\rm \nu^* \rightarrow \ell
W$, and neutral-current decays $\rm \ell^* \rightarrow \ell Z$ and
$\rm \nu^* \rightarrow \nu Z$. The branching fractions in these
different modes depend on the relative values of $f$ and $f'$. As an
example, Table~\ref{tab:bratio} lists these fractions for two mass
values and the the two extreme cases $f = f'$ and $f =
-f'$.  For $f=f'$, the radiative decay is allowed for charged
excited leptons whereas it is forbidden for excited neutrinos.  The
opposite holds for $f=-f'$. The final state topologies and hence the
experimental sensitivity for the two scenarios are therefore very
different.  Table~\ref{tab:finalStates} summarises all the final states for
excited lepton production and decay that are considered in this
Letter.

%
%
\section{Data and Monte Carlo Samples}

The data sample discussed in this letter comprises 216.9 $\rm pb^{-1}$
collected with the L3 detector\cite{l3-detector} at
$\sqrt{s}=202-209\GeV$ with an average centre-of-mass energy of
$206\GeV$.

Monte Carlo samples of pair-produced excited leptons are generated for
a mass of $101 \GeV$, which is close to the expected kinematic
limit. The single production of excited leptons is modeled for masses
of 110, 160 and $201 \GeV$. An interpolation, which includes samples
at 100, 150 and $195 \GeV$ produced at
$\sqrt{s}=192-202\GeV$\cite{L3_192}, allows the estimation of selection
efficiencies in the mass range from 90 to $209 \GeV$. All possible final
states listed in Table~\ref{tab:finalStates} are generated, including both
hadronic and leptonic decays of the W and Z bosons. Differential cross
sections are modeled according to Reference~\citen{hagi} and initial
state radiation is taken into account in cross section calculations.

Standard Model background processes are simulated with several Monte
Carlo generators.  Radiative Bhabha events are generated using
BHWIDE\cite{bhwide} and TEEGG\cite{tee}. KK2F\cite{kk2f} is used for
the $\rm e^+e^- \rightarrow \mu\mu(\gamma)$, $\rm e^+e^- \rightarrow
\tau\tau(\gamma)$ $\rm e^+e^- \rightarrow \nu\nu(\gamma)$ and $\rm
e^+e^- \rightarrow q\bar{q}(\gamma)$ processes. The $\rm e^+e^-
\rightarrow W^+W^-$ process is modeled with KORALW\cite{koralw}, with
the exception of the $\rm q \bar{q} e \nu$ final state, described by
EXCALIBUR\cite{exca} which is also used for the $\rm e^+e^-
\rightarrow\ell\ell\ell\ell$ and $\rm e^+e^-
\rightarrow\ell\ell\nu\nu$ processes. PYTHIA\cite{pythia} is used for
the final states coming from $\rm e^+e^- \rightarrow ZZ$ not covered
by EXCALIBUR. GGG\cite{ggg} describes the $\rm e^+e^-
\rightarrow\gamma\gamma(\gamma)$ process.  The production of hadrons
and leptons in two-photon interactions is described by
PHOJET\cite{phojet} and DIAG36\cite{diag36}, respectively.

The L3 detector response is simulated for all Monte Carlo samples
using the GEANT program\cite{geant}, which includes the
effects of energy loss, multiple scattering and showering in the
detector. Time dependent detector behaviour, as monitored during the
data taking period, is also taken into account.

%
%
\section{Selection Strategy}

The search for charged and neutral excited leptons follows the one
already performed at lower centre-of-mass
energies~\cite{L3_189,L3_192}. Several selections are devised in order
to cover all final states listed in Table~2. All those selections
proceed from the identification of photons, leptons and jets and then
assemble those constituents to single out the particular signature of
each final state.

Photons and electrons are identified in the electromagnetic
calorimeter as clusters within a polar angle with respect to the beam line 
which satisfies $|\cos{\theta}| < 0.95$
and with energy above $1 \GeV$. The shape of the shower in the
crystals must be compatible with a photon or an
electron.  Electrons must be associated to tracks in the central
tracker, while no track is allowed near photon candidates.

Muons are selected from tracks reconstructed in the central or
forward-backward muon spectrometers. These tracks must point to the
interaction vertex. Signals in the time-of-flight system are used to
reject background from cosmic rays. In addition, muon candidates
without a track in the spectrometer are also built starting from energy
depositions in the electromagnetic and hadronic calorimeters consistent with a minimum ionising
particle matched to a track in the central tracker.

Tau leptons are reconstructed from low multiplicity narrow hadronic
jets, or identified through their decay into electrons or muons with
reconstructed missing energy ($\notE$) and momentum.

Jets are reconstructed from charged tracks and energy clusters in the
electromagnetic and hadronic calorimeters. The missing energy and
missing momentum of the event, used to tag events with neutrinos in
the final state, are calculated from all tracks, energy clusters and
additional muons.  The missing momentum is required to point away from
the beam axis in most of the selections, so as to minimise the background
from two-photon interactions and fermion pair production in association
with a high energy initial state photon radiated at low polar angle.

A short summary of the relevant features of the analyses for the pair-
and single-production mechanisms is given in the following two sections.
Details on the different selections are discussed 
elsewhere~\cite{L3_192, Miguel}.

%
%
\section{Pair Production}

The search for pair-produced charged and neutral excited leptons
relies on many selections, aimed to maximise the sensitivity for
excited lepton masses close to the kinematic limit, around
$\sqrt{s}/2$. For these mass values, radiative and charged-current
decays include over 80\% of the decay modes, for any choice of the
couplings. Neutral-current decays are hence not considered.
Table~\ref{tab:selectionsP} summarises the selections used for the
analysis of each final state, as described below.

Radiative decays of charged excited leptons are searched for through
exclusive final states with a pair of identified leptons and two
identified photons. Final states with two photons and missing energy
and momentum are used to tag radiative decays of excited neutrinos.

Four selections are devised to identify the pair of W bosons produced
by the charged-current decays of both charged and neutral excited
leptons. Three selections, collectively denoted as $\rm qq\ell\nu$,
and different according to the lepton flavour, select final states
with two hadronic jets and an isolated high energy lepton. The fourth
selection, $\rm qqqq$, identifies high multiplicity final states, with
four hadronic jets. The $\rm qq\ell\nu$ and $\rm qqqq$ selections are
used for the charged-current decays of charged excited leptons and tau
excited neutrinos. Figure~1a shows the hadronic mass for events
selected by the $\rm qq\ell\nu$ selection. In Figure~1b is shown the
sum of the invariant and recoil masses for two jets associated to
a reconstructed W in the $\rm qqqq$ selection. In the search for
electron and muon excited neutrinos, an additional pair of electrons
and muons is required in addition to the $\rm qq\ell\nu$ and $\rm
qqqq$ selections.

Events where one excited lepton decays radiatively and the other
through the charged-current are investigated with a similar selection for
both charged and neutral excited leptons. Events are selected that
have an isolated lepton, a photon and missing energy. In
addition, the decay products of a W boson are required as either two
hadronic jets or an additional isolated lepton. Tighter requirements on the amount 
of missing energy are applied for the case of neutral excited leptons. Figures~1c and 1d show
the masses of the excited electron and muon candidates, reconstructed
as the photon-lepton invariant mass.

%
%
\section{Single Production}

The search for singly produced excited leptons complements the search
through the pair-production mechanism and gives access to the mass
range from $\sqrt{s}/2$ up to $\sqrt{s}$.  To retain the highest
efficiency, all decay modes are investigated: radiative,
charged-current and neutral-current. Table~\ref{tab:selectionsS} lists
the association between all final states and the corresponding
selections which are summarised below.

In the search for charged excited leptons decaying radiatively, final
states with two leptons and a photon are selected. High sensitivity to
the excited lepton mass is achieved via the invariant mass of a
detected lepton and the photon, as presented in Figures~2a, 2b and
2c. Events with one photon and large missing energy and momentum are
used to search for the possible production of excited
neutrinos. Figure 2d presents the normalised energy spectrum of the
selected photons.

For the charged-current decay of charged and neutral excited leptons,
the $\rm q q \ell \nu$ selection of the pair-production searches is used.  Figure~3
presents the distribution used for the reconstruction of the mass of
the excited lepton candidates, namely the lepton recoil mass in the
case of charged excited leptons and the invariant mass of the lepton
and the detected jets in the case of excited neutrinos with a hadronic
decay of the W boson. To retain the highest efficiency, two additional
selections are devised. They identify events with hadronic activity
and missing energy, $\rm qq \notE$, as well as leptons and missing
energy, $\rm \ell\ell \notE$. These selections complement the $\rm q q
\ell \nu$ selection in the case in which the lepton is not detected or
the W boson decays into leptons.

Neutral-current decays of charged excited leptons are searched for
with the same selections used for the charged-current case,
supplementing the  $\rm \ell\ell \notE$ by the requirement that
the visible mass of the two leptons is close to the Z boson mass
The
neutral-current decay of excited neutrinos produces a Z boson
and missing energy, searched for with the $\rm qq \notE$ 
and  $\rm \ell\ell \notE$ selections. The latter requires a
visible mass of the two leptons close to the mass of the Z boson.

%
%
\section{Results}

Tables~\ref{tab:selectionsP} and~\ref{tab:selectionsS} list the number
of events observed in the data by each selection together with the
Monte Carlo background expectations and the signal efficiencies. No
evidence for the production of excited leptons is observed in any
final state. 

Systematic uncertainties affect the results in
Tables~\ref{tab:selectionsP} and~\ref{tab:selectionsS}. An uncertainty
of $1.5-2.5\%$, depending on the selection, is associated to the
background estimate. This includes the uncertainties on the cross
sections of background processes,  limited Monte Carlo statistics,
detector simulation and the selection procedure. The limited Monte
Carlo statistics and the detector simulation also affect the estimation
of the signal efficiency. Depending on the selection, the systematic
uncertainty from this source is around 2.5\%, which also covers the
accuracy of the efficiency interpolation for different excited lepton
masses.

The absence of excited leptons in the data sample is expressed by
means of upper limits on their masses and couplings.  In the
derivation of these limits, the data discussed in this Letter are
considered at their luminosity averaged $\sqrt{s}$ of $206\GeV$. Absolute limits
on the masses of charged and neutral excited leptons are derived from the pair-production process, whereas
limits on the effective couplings as a function of the masses are
derived from the single-production study. The limits discussed in the
following also include the results from previous searches at $\sqrt{s}
= 189 \GeV$\cite{L3_189} and $\sqrt{s} = 192-202
\GeV$\cite{L3_192}. All limits take systematic uncertainties into
account and are reported for the two scenarios, $f=f'$ and
$f=-f'$, which reflect the different dominant branching ratio and
consequent final state topologies. For the pair-production process, the numbers of observed
and expected events, together with the signal efficiencies are
translated into upper limits on the cross section for the production
of excited leptons.  A scan is performed for all possible values of
the ratio $f/f'$, and the decay fractions of charged excited leptons
and excited neutrinos are computed for the radiative, charged-current
and mixed decay modes.  Signal cross sections are calculated and the
corresponding mass limits are derived, as presented in
Table~\ref{tab:limits}. More stringent limits are obtained for
channels with low background. As an example, the limits on charged
excited leptons in the $f=f'$ scenario and excited neutrinos in the
$f=-f'$ scenario benefit from a large branching ratio in the clean
radiative decay channel.  In addition to those corresponding to the
$f=f'$ and $f=-f'$ scenarios, limits are also given for all excited
leptons which correspond to the lowest values obtained in the scan
over $f/f'$, and hence valid for any choice of the couplings.

In the case of single-production searches, an upper limit on the cross
section is obtained as a function of the excited lepton
mass. Different mass values are investigated by means of the
distributions of the variables presented in Figures~\ref{fig:rad}
and~\ref{fig:weak}. A linear interpolation of the detection
efficiencies as a function of the excited lepton mass is used.
These cross section limits are translated in the
upper limits on the ratios $|f|/\Lambda$ and $|f'|/\Lambda$ shown in
Figure~\ref{fig:limites}.  The edge of the curves at low mass
indicates the lower mass limit derived from pair-production searches,
whereas the rise at high mass reflects the decrease of the expected
signal cross section and therefore of the experimental sensitivity.
The limits corresponding to charged excited leptons in the $f=f'$
scenario and excited neutrinos in the $f=-f'$ scenario are derived
mainly from the radiative decay searches, whose clean final states
have a high signal sensitivity. These limits are more stringent than
those obtained in the complementary scenarios in which the radiative
decays are forbidden.  The limits corresponding to excited leptons of
the first generation are significantly tighter due to their higher
cross section resulting from the $t$-channel contribution.

In conclusion, no evidence for charged and neutral excited leptons of
any flavour is found in the LEP data, and lower mass limits as high as
$101.5 \GeV$ are derived for any value of the excited lepton
couplings. Upper limits on $|f|/\Lambda$ and $|f'|/\Lambda$, ranging
from $10^{-1}$ to $10^{-4} \GeV^{-1}$ according to the excited lepton
flavour and mass, are set in the mass range from 100 to $200\GeV$.

%
%

%
%
  \newpage
  \typeout{   }     
\typeout{Using author list for paper 261 -  }
\typeout{$Modified: Jul 15 2001 by smele $}
\typeout{!!!!  This should only be used with document option a4p!!!!}
\typeout{   }
%
%
%
%
%
%

\newcount\tutecount  \tutecount=0
\def\tutenum#1{\global\advance\tutecount by 1 \xdef#1{\the\tutecount}}
\def\tute#1{$^{#1}$}
\tutenum\aachen            
\tutenum\nikhef            
\tutenum\mich              
\tutenum\lapp              
\tutenum\basel             
\tutenum\lsu               
\tutenum\beijing           
\tutenum\bologna           
\tutenum\tata              
\tutenum\ne                
\tutenum\bucharest         
\tutenum\budapest          
\tutenum\mit               
\tutenum\panjab            
\tutenum\debrecen          
\tutenum\dublin            
\tutenum\florence          
\tutenum\cern              
\tutenum\wl                
\tutenum\geneva            
\tutenum\hefei             
\tutenum\lausanne          
\tutenum\lyon              
\tutenum\madrid            
\tutenum\florida           
\tutenum\milan             
\tutenum\moscow            
\tutenum\naples            
\tutenum\cyprus            
\tutenum\nymegen           
\tutenum\caltech           
\tutenum\perugia           
\tutenum\peters            
\tutenum\cmu               
\tutenum\potenza           
\tutenum\prince            
\tutenum\riverside         
\tutenum\rome              
\tutenum\salerno           
\tutenum\ucsd              
\tutenum\sofia             
\tutenum\korea             
\tutenum\purdue            
\tutenum\psinst            
\tutenum\zeuthen           
\tutenum\eth               
\tutenum\hamburg           
\tutenum\taiwan            
\tutenum\tsinghua          

{
\parskip=0pt
\noindent
{\bf The L3 Collaboration:}
\ifx\selectfont\undefined
 \baselineskip=10.8pt
 \baselineskip\baselinestretch\baselineskip
 \normalbaselineskip\baselineskip
 \ixpt
\else
 \fontsize{9}{10.8pt}\selectfont
\fi
\medskip
\tolerance=10000
\hbadness=5000
\raggedright
\hsize=162truemm\hoffset=0mm
\def\r{\rlap,}
\noindent

P.Achard\r\tute\geneva\ 
O.Adriani\r\tute{\florence}\ 
M.Aguilar-Benitez\r\tute\madrid\ 
J.Alcaraz\r\tute{\madrid}\ 
G.Alemanni\r\tute\lausanne\
J.Allaby\r\tute\cern\
A.Aloisio\r\tute\naples\ 
M.G.Alviggi\r\tute\naples\
H.Anderhub\r\tute\eth\ 
V.P.Andreev\r\tute{\lsu,\peters}\
% ...
\newpage
\begin{list}{A}{\itemsep=0pt plus 0pt minus 0pt\parsep=0pt plus 0pt minus 0pt
                \topsep=0pt plus 0pt minus 0pt}
\item[\aachen]
 III. Physikalisches Institut, RWTH, D-52056 Aachen, Germany$^{\S}$
\item[\nikhef] National Institute for High Energy Physics, NIKHEF, 
     and University of Amsterdam, NL-1009 DB Amsterdam, The Netherlands
\item[\mich] University of Michigan, Ann Arbor, MI 48109, USA
\item[\lapp] Laboratoire d'Annecy-le-Vieux de Physique des Particules, 
     LAPP,IN2P3-CNRS, BP 110, F-74941 Annecy-le-Vieux CEDEX, France
\item[\basel] Institute of Physics, University of Basel, CH-4056 Basel,
     Switzerland
\item[\lsu] Louisiana State University, Baton Rouge, LA 70803, USA
\item[\beijing] Institute of High Energy Physics, IHEP, 
  100039 Beijing, China$^{\triangle}$ 
\item[\bologna] University of Bologna and INFN-Sezione di Bologna, 
     I-40126 Bologna, Italy
\item[\tata] Tata Institute of Fundamental Research, Mumbai (Bombay) 400 005, India
\item[\ne] Northeastern University, Boston, MA 02115, USA
\item[\bucharest] Institute of Atomic Physics and University of Bucharest,
     R-76900 Bucharest, Romania
\item[\budapest] Central Research Institute for Physics of the 
     Hungarian Academy of Sciences, H-1525 Budapest 114, Hungary$^{\ddag}$
\item[\mit] Massachusetts Institute of Technology, Cambridge, MA 02139, USA
\item[\panjab] Panjab University, Chandigarh 160 014, India.
\item[\debrecen] KLTE-ATOMKI, H-4010 Debrecen, Hungary$^\P$
\item[\dublin] Department of Experimental Physics,
  University College Dublin, Belfield, Dublin 4, Ireland
\item[\florence] INFN Sezione di Firenze and University of Florence, 
     I-50125 Florence, Italy
\item[\cern] European Laboratory for Particle Physics, CERN, 
     CH-1211 Geneva 23, Switzerland
\item[\wl] World Laboratory, FBLJA  Project, CH-1211 Geneva 23, Switzerland
\item[\geneva] University of Geneva, CH-1211 Geneva 4, Switzerland
\item[\hefei] Chinese University of Science and Technology, USTC,
      Hefei, Anhui 230 029, China$^{\triangle}$
\item[\lausanne] University of Lausanne, CH-1015 Lausanne, Switzerland
\item[\lyon] Institut de Physique Nucl\'eaire de Lyon, 
     IN2P3-CNRS,Universit\'e Claude Bernard, 
     F-69622 Villeurbanne, France
\item[\madrid] Centro de Investigaciones Energ{\'e}ticas, 
     Medioambientales y Tecnol\'ogicas, CIEMAT, E-28040 Madrid,
     Spain${\flat}$ 
\item[\florida] Florida Institute of Technology, Melbourne, FL 32901, USA
\item[\milan] INFN-Sezione di Milano, I-20133 Milan, Italy
\item[\moscow] Institute of Theoretical and Experimental Physics, ITEP, 
     Moscow, Russia
\item[\naples] INFN-Sezione di Napoli and University of Naples, 
     I-80125 Naples, Italy
\item[\cyprus] Department of Physics, University of Cyprus,
     Nicosia, Cyprus
\item[\nymegen] University of Nijmegen and NIKHEF, 
     NL-6525 ED Nijmegen, The Netherlands
\item[\caltech] California Institute of Technology, Pasadena, CA 91125, USA
\item[\perugia] INFN-Sezione di Perugia and Universit\`a Degli 
     Studi di Perugia, I-06100 Perugia, Italy   
\item[\peters] Nuclear Physics Institute, St. Petersburg, Russia
\item[\cmu] Carnegie Mellon University, Pittsburgh, PA 15213, USA
\item[\potenza] INFN-Sezione di Napoli and University of Potenza, 
     I-85100 Potenza, Italy
\item[\prince] Princeton University, Princeton, NJ 08544, USA
\item[\riverside] University of Californa, Riverside, CA 92521, USA
\item[\rome] INFN-Sezione di Roma and University of Rome, ``La Sapienza",
     I-00185 Rome, Italy
\item[\salerno] University and INFN, Salerno, I-84100 Salerno, Italy
\item[\ucsd] University of California, San Diego, CA 92093, USA
\item[\sofia] Bulgarian Academy of Sciences, Central Lab.~of 
     Mechatronics and Instrumentation, BU-1113 Sofia, Bulgaria
\item[\korea]  The Center for High Energy Physics, 
     Kyungpook National University, 702-701 Taegu, Republic of Korea
\item[\purdue] Purdue University, West Lafayette, IN 47907, USA
\item[\psinst] Paul Scherrer Institut, PSI, CH-5232 Villigen, Switzerland
\item[\zeuthen] DESY, D-15738 Zeuthen, Germany
\item[\eth] Eidgen\"ossische Technische Hochschule, ETH Z\"urich,
     CH-8093 Z\"urich, Switzerland
\item[\hamburg] University of Hamburg, D-22761 Hamburg, Germany
\item[\taiwan] National Central University, Chung-Li, Taiwan, China
\item[\tsinghua] Department of Physics, National Tsing Hua University,
      Taiwan, China
\item[\S]  Supported by the German Bundesministerium 
        f\"ur Bildung, Wissenschaft, Forschung und Technologie
\item[\ddag] Supported by the Hungarian OTKA fund under contract
numbers T019181, F023259 and T037350.
\item[\P] Also supported by the Hungarian OTKA fund under contract
  number T026178.
\item[$\flat$] Supported also by the Comisi\'on Interministerial de Ciencia y 
        Tecnolog{\'\i}a.
\item[$\sharp$] Also supported by CONICET and Universidad Nacional de La Plata,
        CC 67, 1900 La Plata, Argentina.
\item[$\triangle$] Supported by the National Natural Science
  Foundation of China.
\end{list}
}
\vfill


%

\begin{table}[th]
\begin{center}
\begin{tabular}{|rcl|c|c|c|c|}\hline
  & &    &  \multicolumn{4}{|c|}{Branching Ratios} \\
\cline{4-7}
\multicolumn{3}{|c|}{Decay}    &  \multicolumn{2}{|c|}{$M=102\GeV$} &
          \multicolumn{2}{|c|}{$M=200\GeV$}    \\
         \cline{4-7}
\multicolumn{3}{|c|}{Channel}  & $f=f'$ & $f=-f'$ & $f=f'$ & $f=-f'$  \\
\hline
$\rm \ell^*$ & $ \rightarrow $ & $\rm \ell \gamma$ &
          70\%   &    --           &  35\%   &   --       \\
$\rm \ell^* $ & $\rightarrow $ & $\rm \nu W      $ &
          28\%   &   83\%          &  55\%   &  63\%      \\
$\rm \ell^* $ & $\rightarrow $ & $\rm \ell Z     $ &
          ~2\%   &   17\%          &  10\%   &  37\%      \\
\hline
$\rm \nu^*  $ & $\rightarrow $ & $\rm \nu  \gamma$ &
           --    &   70\%          &   --    &  35\%      \\
$\rm \nu^*  $ & $\rightarrow $ & $\rm \ell W     $ &
          83\%   &   28\%          &  63\%   &  55\%      \\
$\rm \nu^*  $ & $\rightarrow $ & $\rm \nu  Z     $ &
          17\%   &   ~2\%          &  37\%   &  10\%      \\
\hline
\end{tabular}
\icaption{Predicted branching ratios for charged and neutral
      excited lepton decays, for different choices of masses ($M$)
              and couplings.
    \label{tab:bratio}}
  \end{center}
\end{table}
\begin{table}[th]
\begin{center}
\begin{tabular}{|c|c|c|c|c|}
\hline
Decay & \multicolumn{2}{c|}{Pair production} 
 & \multicolumn{2}{c|}{Single production} \\
\cline{2-5}
mode & $ \rm e^+ e^- \rightarrow \ell^* \ell^*$
& $ \rm e^+ e^- \rightarrow \nu^* \nu^*$ 
& $ \rm e^+ e^- \rightarrow \ell  \ell^*$
& $ \rm e^+ e^- \rightarrow \nu  \nu^*$ \\

\hline 
\rule{0pt}{13pt} 
Radiative 
& $\rm \ell^* \ell^* \rightarrow\rm \ell \ell \gamma \gamma$ 
& $\rm \nu^*\nu^* \rightarrow\rm \nu \nu \gamma \gamma$
& $\rm \ell \ell^* \rightarrow\rm \ell \ell \gamma$
& $\rm \nu \nu^* \rightarrow\rm \nu \nu \gamma$   
\\
   
\hline 
\rule{0pt}{13pt} 
Charged-current 
& $\rm \ell^* \ell^* \rightarrow\rm \nu \nu W W$
& $\rm \nu_{\ell}^* \nu_{\ell}^* \rightarrow\rm \ell \ell W W$
& $\rm \nu_{\ell} \nu_{\ell}^* \rightarrow\rm \nu_{\ell} \ell W$ 
& $\rm \ell \ell^* \rightarrow\rm \ell \nu_{\ell} W$             
\\

\hline
\rule{0pt}{13pt} 
Neutral-current 
& --
& --
& $\rm \ell \ell^* \rightarrow\rm \ell \ell Z$
& $\rm \nu \nu^* \rightarrow\rm \nu \nu Z$
\\

\hline
\rule{0pt}{13pt} 
Mixed 
& $\rm \ell^* \ell^* \rightarrow\rm \ell \gamma \nu W $
& $\rm \nu_{\ell}^* \nu_{\ell}^* \rightarrow\rm \nu \gamma \ell W $
& --
& --
\\
\hline

\end{tabular}
\icaption{The final states for excited lepton production considered in
  this Letter, where ${\ell}$ runs on the lepton flavour: e, $\mu$ and $\tau$.
\label{tab:finalStates}}
\end{center}
\end{table}

\begin{table}[th]
\begin{center}
\begin{tabular}{|c|c|c|c|c|}
\hline
\multicolumn{5}{|c|}{Pair production}  \\
\hline
Signal & Selection & $N_D$ & $N_B$ & $\varepsilon$ (\%) \\
\hline 
\rule{0pt}{13pt} 
$\rm e^* e^* \rightarrow \rm e e \gamma \gamma $                
& Leptons
& 0   &    0.1  &   45            \\
\rule{0pt}{13pt} 
$\rm \mu^* \mu^* \rightarrow\rm \mu \mu \gamma \gamma$ 
& and
& 0    &    0.3  &   44           \\
\rule{0pt}{13pt} 
$\rm \tau^* \tau^* \rightarrow\rm \tau \tau \gamma \gamma$ 
& photons
& 0    &    0.4  &   40           \\
\hline
\rule{0pt}{13pt} 
$\rm \ell^* \ell^* \rightarrow\rm \nu \nu W W$
& $\rm qqqq$ or $\rm qq\ell\nu$
& 3012  & 2974   &   66            \\
\hline
\rule{0pt}{13pt} 
$\rm e^* e^* \rightarrow\rm e \gamma \nu W $
& $\rm e + \gamma +\notE + [qq\,\, or\,\, \ell]$
&  9   &   10    &   56            \\
\rule{0pt}{13pt} 
$\rm \mu^* \mu^* \rightarrow\rm \mu \gamma \nu W $
& $\rm \mu + \gamma +\notE + [qq\,\, or\,\, \ell]$
&  6    &   5     &   49           \\
\rule{0pt}{13pt}
$\rm \tau^* \tau^* \rightarrow\rm \tau \gamma \nu W $
&$\rm jet + \gamma +\notE + [qq\,\, or\,\, \ell]$
&  45   &   43    &   35           \\
\hline
\rule{0pt}{13pt} 
$\rm \nu^*\nu^* \rightarrow\rm \nu \nu \gamma \gamma$
& $\gamma\gamma + \notE$
&  2    &    1.9  &   45            \\
\hline
\rule{0pt}{13pt} 
$\rm \nu_{e}^* \nu_{e}^* \rightarrow \rm e e W W$
&  $\rm ee + [qqqq\,\, or\,\, qq\ell\nu]$
&  1    &    0.2  &   15            \\
\rule{0pt}{13pt} 
$\rm \nu_{\mu}^* \nu_{\mu}^* \rightarrow\rm \mu \mu W W$
& $\rm \mu\mu + [qqqq\,\, or\,\, qq\ell\nu]$
&  2    &    0.6  &   19            \\
\rule{0pt}{13pt} 
$\rm \nu_{\tau}^* \nu_{\tau}^* \rightarrow\rm \tau \tau W W$
&$\rm qqqq\,\, or\,\, qq\ell\nu$
& 3012  & 2974   &   70            \\
\hline
\rule{0pt}{13pt} 
$\rm \nu_{e}^* \nu_{e}^* \rightarrow\rm \nu \gamma e W$
&$\rm e + \gamma +\notE + [qq\,\, or\,\, \ell]$
&       &         &   35            \\
\rule{0pt}{13pt} 
$\rm \nu_{\mu}^* \nu_{\mu}^* \rightarrow\rm \nu \gamma \mu W $
& $\rm \mu + \gamma +\notE + [qq\,\, or\,\, \ell]$
& 10   &   11    &   29            \\
\rule{0pt}{13pt}
$\rm \nu_{\tau}^* \nu_{\tau}^* \rightarrow\rm \nu \gamma \tau W $
&$\rm jet + \gamma +\notE + [qq\,\, or\,\, \ell]$
&       &         &   23            \\
\hline
\end{tabular}
\icaption{Selections used in the search for pair-produced excited
  leptons and the corresponding numbers of observed events, $N_D$, expected
  background, $N_B$, and selection efficiency, $\varepsilon$, for an excited 
  lepton mass of $101\GeV$. A tighter cut on the missing energy is
  applied in the $\rm \nu^* \nu^* \rightarrow\rm \nu \gamma \ell W$
  selection with respect to the $\rm \ell^* \ell^* \rightarrow\rm \ell
  \gamma \nu W$ one.
\label{tab:selectionsP}}
\end{center}
\end{table}


\begin{table}[th]
\begin{center}
\begin{tabular}{|c|c|c|c|c|}

\hline
\multicolumn{5}{|c|}{Single Production}  \\
\hline
Signal & Selection & $N_D$ & $N_B$ & $\varepsilon$ (\%) \\
\hline
\rule{0pt}{13pt} 
$\rm e e^* \rightarrow\rm e e \gamma$        
& Leptons
& 672   &  737    &   54,\,65,\,65 \\
\rule{0pt}{13pt} 
$\rm \mu \mu^* \rightarrow\rm \mu \mu \gamma$   
& and a 
& 55    &   66    &   62,\,63,\,60           \\
\rule{0pt}{13pt} 
$\rm \tau \tau^* \rightarrow\rm \tau \tau \gamma$   
& photon
& 48    &   60    &   42,\,45,\,43 \\
\hline
\rule{0pt}{13pt} 
$\rm e e^* \rightarrow\rm e \nu_e W$                
& $\rm qq\ell\nu$ or
& 749  &  773    &   65,\,64,\,61           \\
\rule{0pt}{13pt} 
$\rm \mu \mu^* \rightarrow\rm \mu \nu_{\mu} W$             
& $\rm q q \notE$ or
& 652   &  675    &   60,\,64,\,62             \\
\rule{0pt}{13pt} 
$\rm \tau \tau^* \rightarrow\rm \tau \nu_{\tau} W$
&  $\rm \ell \ell  \notE$
& 1287  & 1307    &   59,\,48,\,50            \\
\hline
\rule{0pt}{13pt} 
$\rm e e^* \rightarrow\rm e e Z$                    
&  $\rm qq\ell\nu$ or
& 673  &  698    &   65,\,37,\,63             \\
\rule{0pt}{13pt} 
$\rm \mu \mu^* \rightarrow\rm \mu \mu Z$            
&  $\rm q q \notE$ or
& 568   &  596    &   58,\,41,\,59            \\
\rule{0pt}{13pt} 
$\rm \tau \tau^* \rightarrow\rm \tau \tau Z$        
&  $\rm \ell \ell  \notE$
& 1217  & 1236    &   32,\,22,\,37 \\
\hline
\rule{0pt}{13pt} 
$\rm \nu \nu^* \rightarrow\rm \nu \nu \gamma$   
& $\gamma +\notE$
& 176   &  194    &   60,\,67,\,70            \\
\hline
\rule{0pt}{13pt} 
$\rm \nu_{e} \nu_{e}^* \rightarrow\rm \nu_{e} e W$  
& $\rm qq\ell\nu$ or
& 749   &  773    &   61,\,68,\,67            \\
\rule{0pt}{13pt} 
$\rm \nu_{\mu} \nu_{\mu}^* \rightarrow\rm \nu_{\mu} \mu W$ 
&  $\rm q q \notE$ or
& 652   &  675    &   63,\,63,\,60            \\
\rule{0pt}{13pt} 
$\rm \nu_{\tau} \nu_{\tau}^* \rightarrow\rm \nu_{\tau} \tau W$
&  $\rm \ell \ell  \notE$
& 1287  & 1307    &   46,\,57,\,60            \\
\hline
\rule{0pt}{13pt} 
$\rm \nu \nu^* \rightarrow\rm \nu \nu Z$            
&  $\rm q q \notE$ or  $\rm \ell \ell  \notE$
& 343  &  350    &  18,\,15,\,40            \\
\hline

\end{tabular}
\icaption{Selections used in the search for singly-produced excited
  leptons and corresponding numbers of observed events, $N_D$, expected
  background, $N_B$, and selection efficiency, $\varepsilon$, for excited 
  lepton masses of 110, 160 and $201\GeV$. The  $\rm \ell \ell  \notE$
  selections used for the neutral-current decays are tighter than
  those used for the charged-current decays.
\label{tab:selectionsS}}
\end{center}
\end{table}


\begin{table}[th]
\begin{center}
\begin{tabular}{|c|c|c|c|}
\hline
    Excited           &  \multicolumn{3}{|c|}{95\% CL Mass Limit (\GeV)} \\
    \cline{2-4}
    Lepton            & $f=f'$ & $f=-f'$& Any Coupling \\
    \hline
    $\rm \e^*         $ & 102.8  &  ~96.6  & ~96.5 \\
    $\rm \mu^*        $ & 102.8  &  ~96.6  & ~96.6 \\
    $\rm \tau^*       $ & 102.8  &  ~96.6  & ~95.6 \\
    \hline
    $\rm \nu^*_e      $ & 101.7  & 102.6  & 101.5 \\
    $\rm \nu^*_{\mu}  $ & 101.8  &  102.6  & 101.4 \\
    $\rm \nu^*_{\tau} $ & ~92.9  &  102.6  & ~91.3 \\
    \hline
\end{tabular}
\icaption{Lower mass limits at 95\%
          confidence level for charged and neutral excited lepton as
          obtained from pair-production searches.
\label{tab:limits}}
\end{center}
\end{table}

\clearpage
%
%

\begin{figure}[htb]
  \begin{center}
    \includegraphics[width=0.49\textwidth]{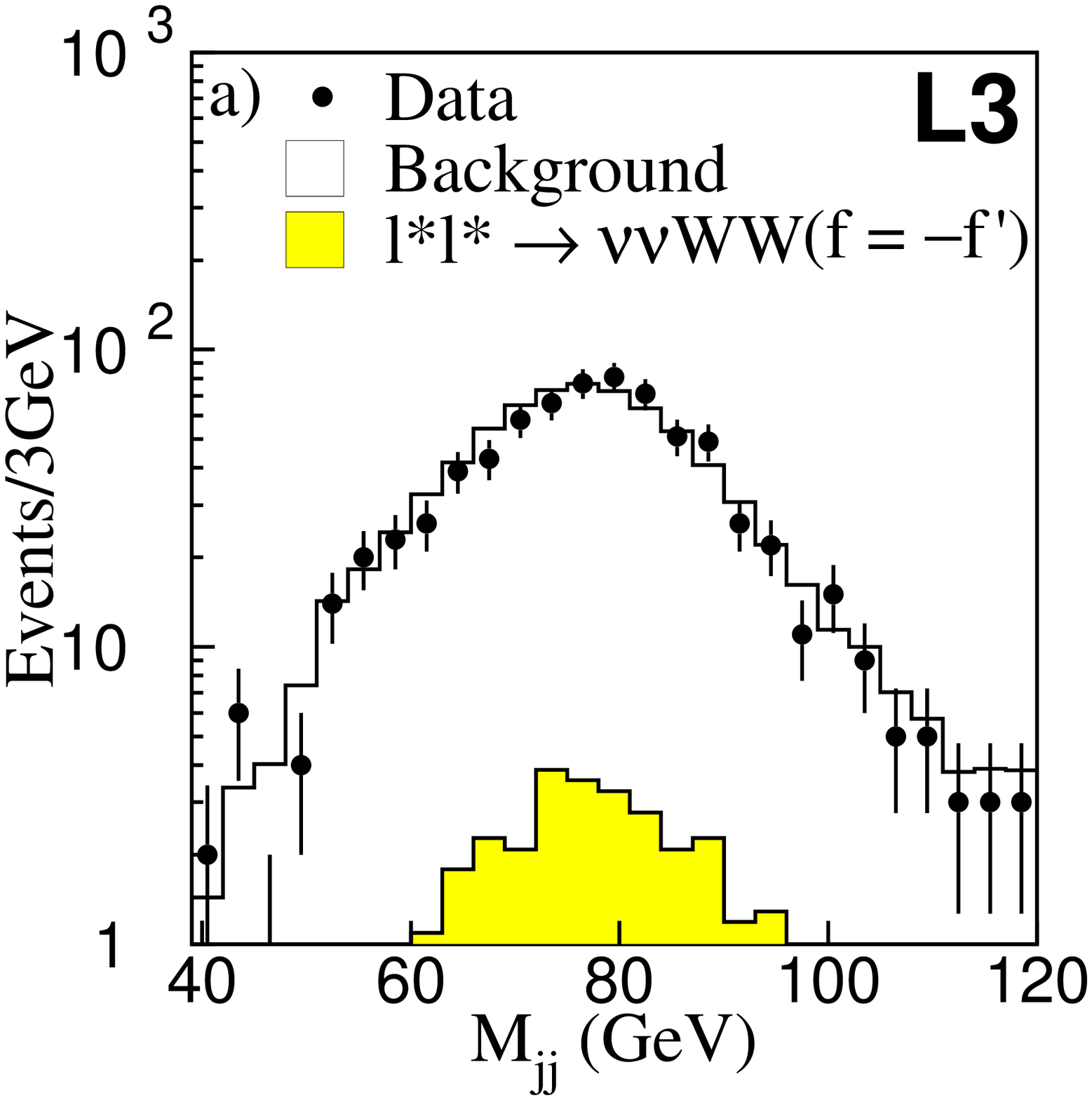}
    \includegraphics[width=0.49\textwidth]{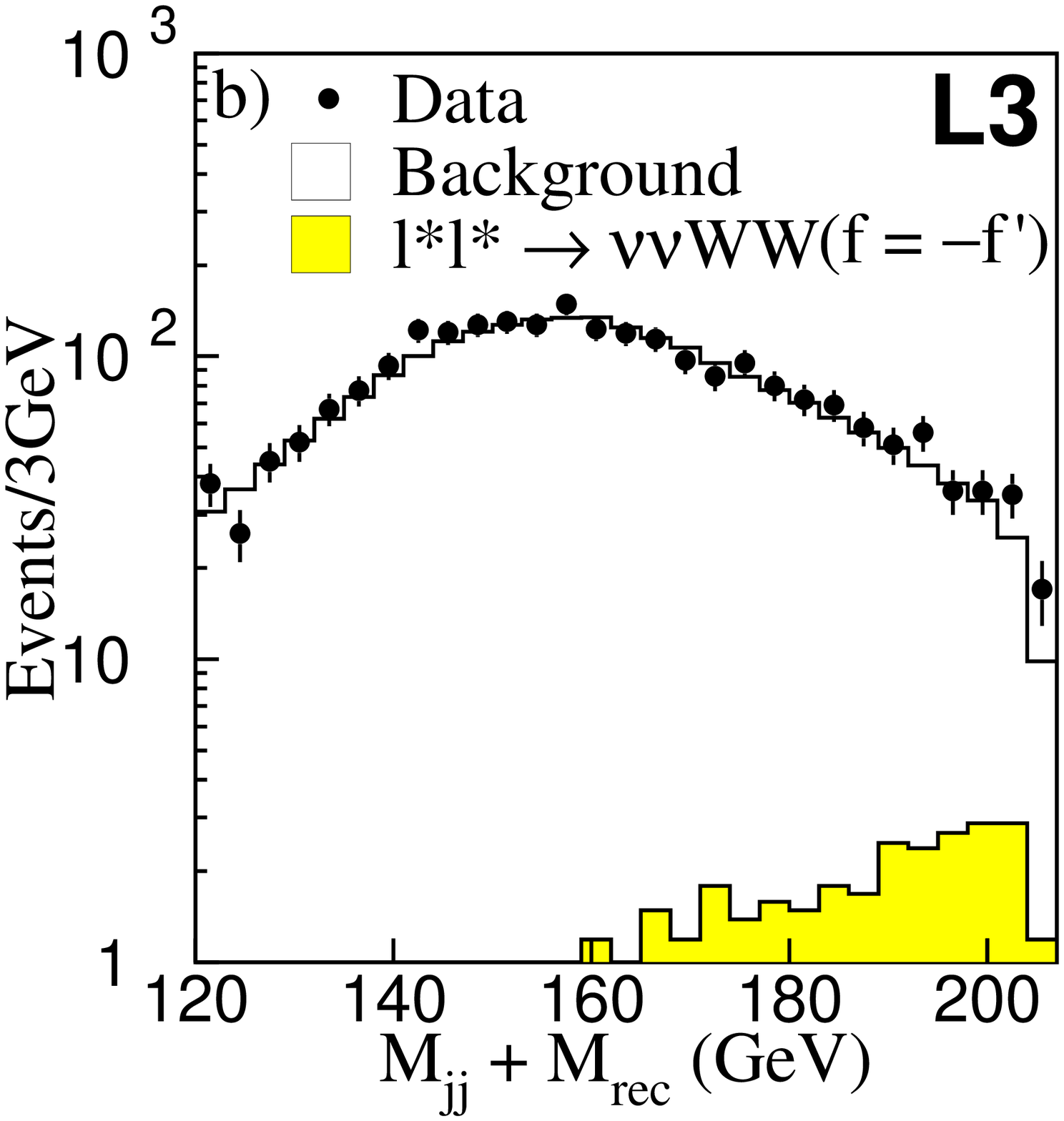}
    \includegraphics[width=0.49\textwidth]{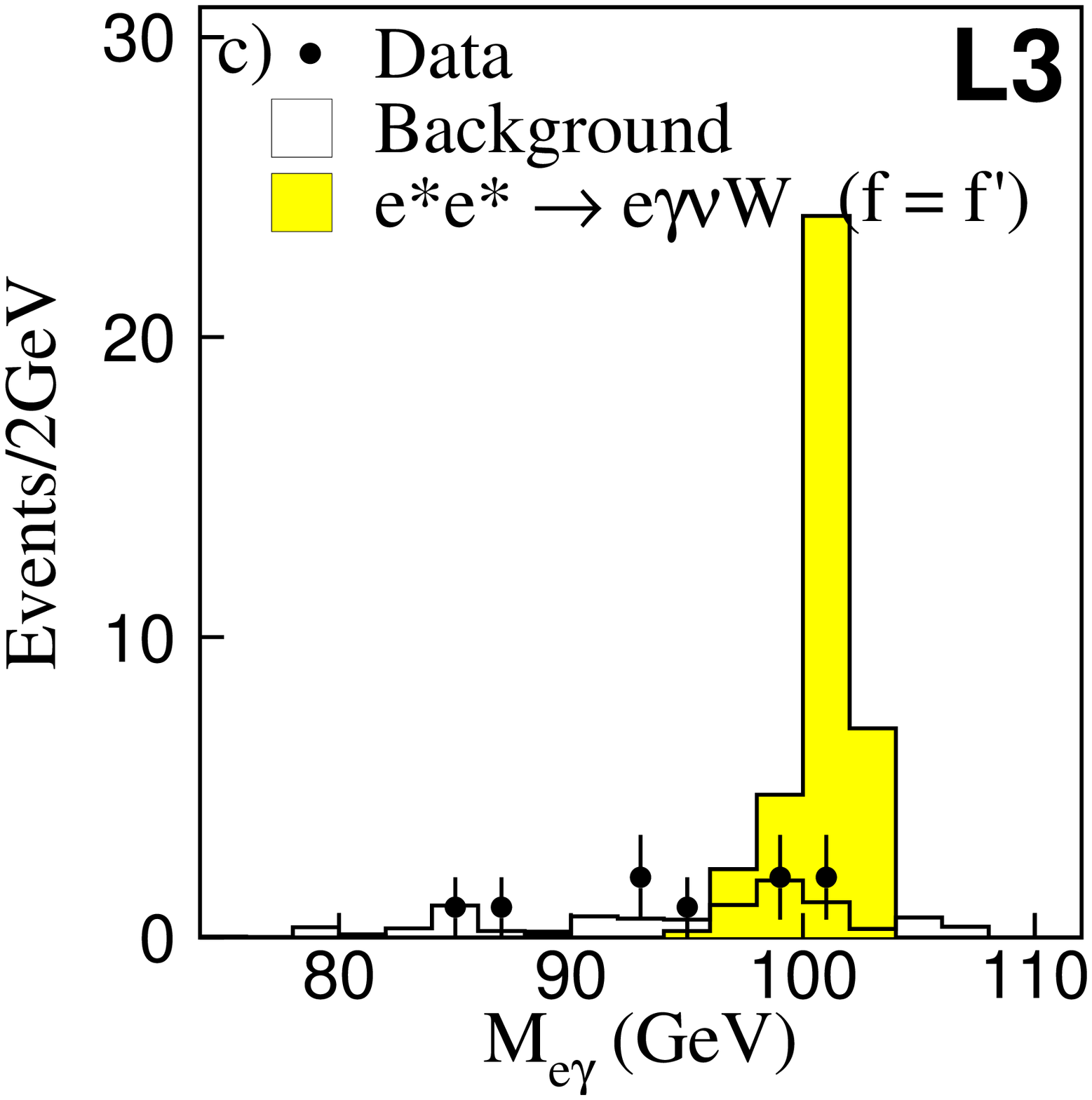}
    \includegraphics[width=0.49\textwidth]{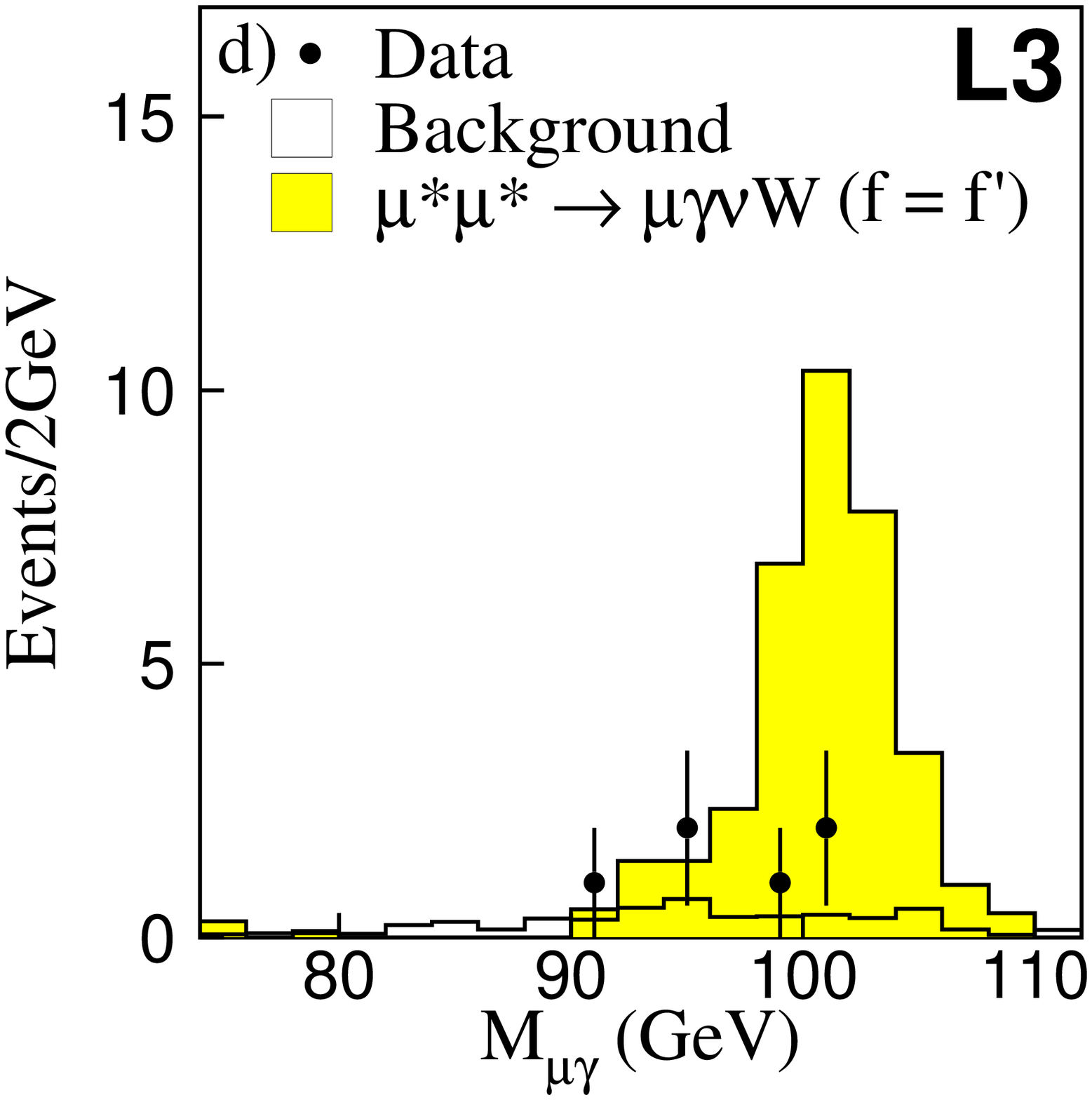}
  \end{center}
  \icaption{Distributions of a) the hadronic invariant mass, $M_{jj}$, in the 
            $\rm qq\ell\nu$ selection;
            b)  the sum of  $M_{jj}$ and the recoil
            mass, $M_{rec}$, for two jets in the $\rm qqqq$ selection;
            c) the electron photon invariant mass in the 
            $\rm e\gamma\nu W$ selection and 
            d) the muon photon invariant mass in the
            $\rm \mu\gamma\nu W$ selection.
            The expected signal for excited leptons produced in pairs with a 
            mass of $101 \GeV$ is shown together with data and 
            Standard Model background, which is dominated by
            charged-current four-fermion production in a) and b) and
            by fermion pair-production with initial state radiation in
            c) and d).
  \label{fig:pair}}
\end{figure}

\begin{figure}[htb]
  \begin{center}
    \includegraphics[width=0.49\textwidth]{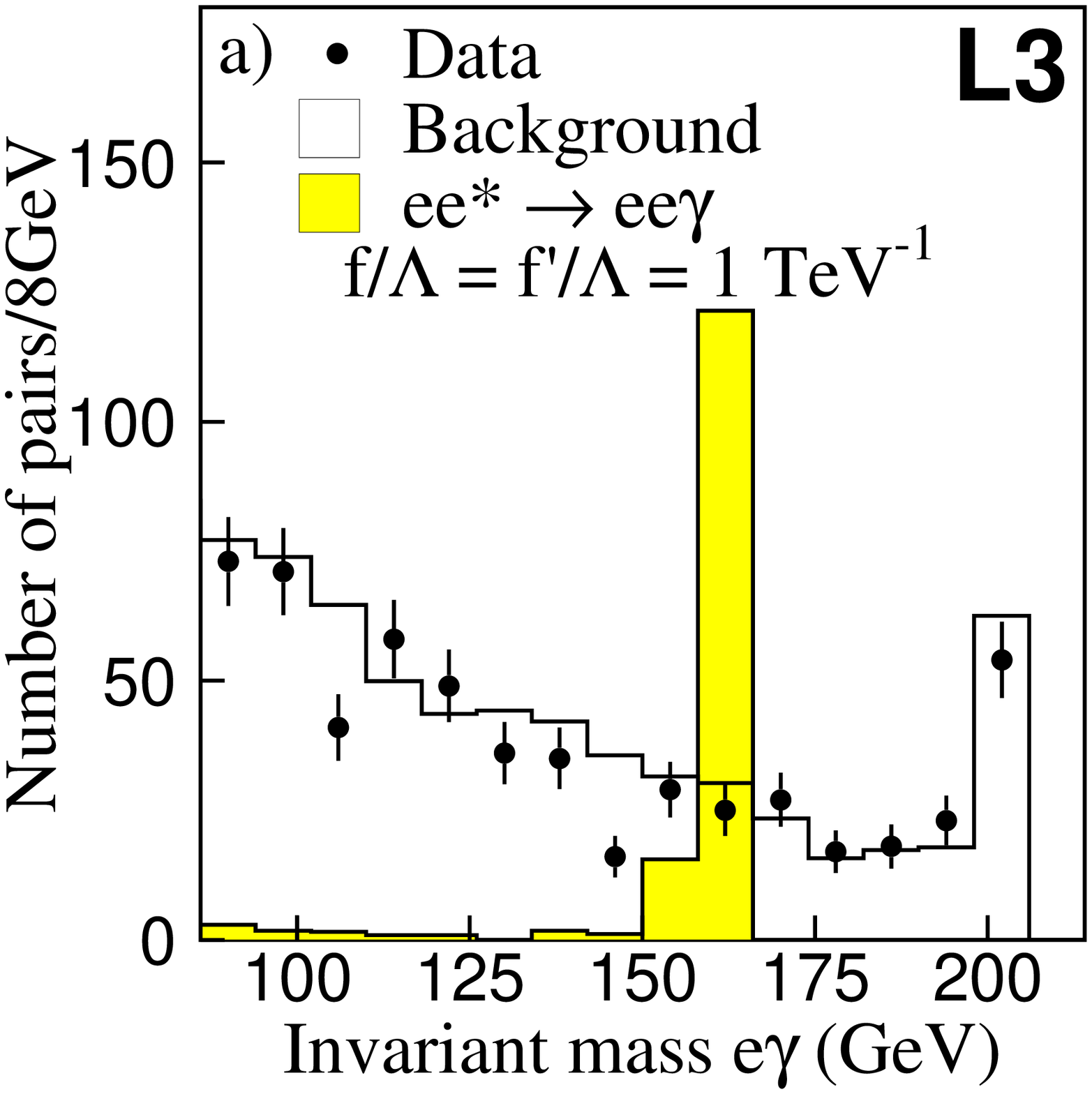}
    \includegraphics[width=0.49\textwidth]{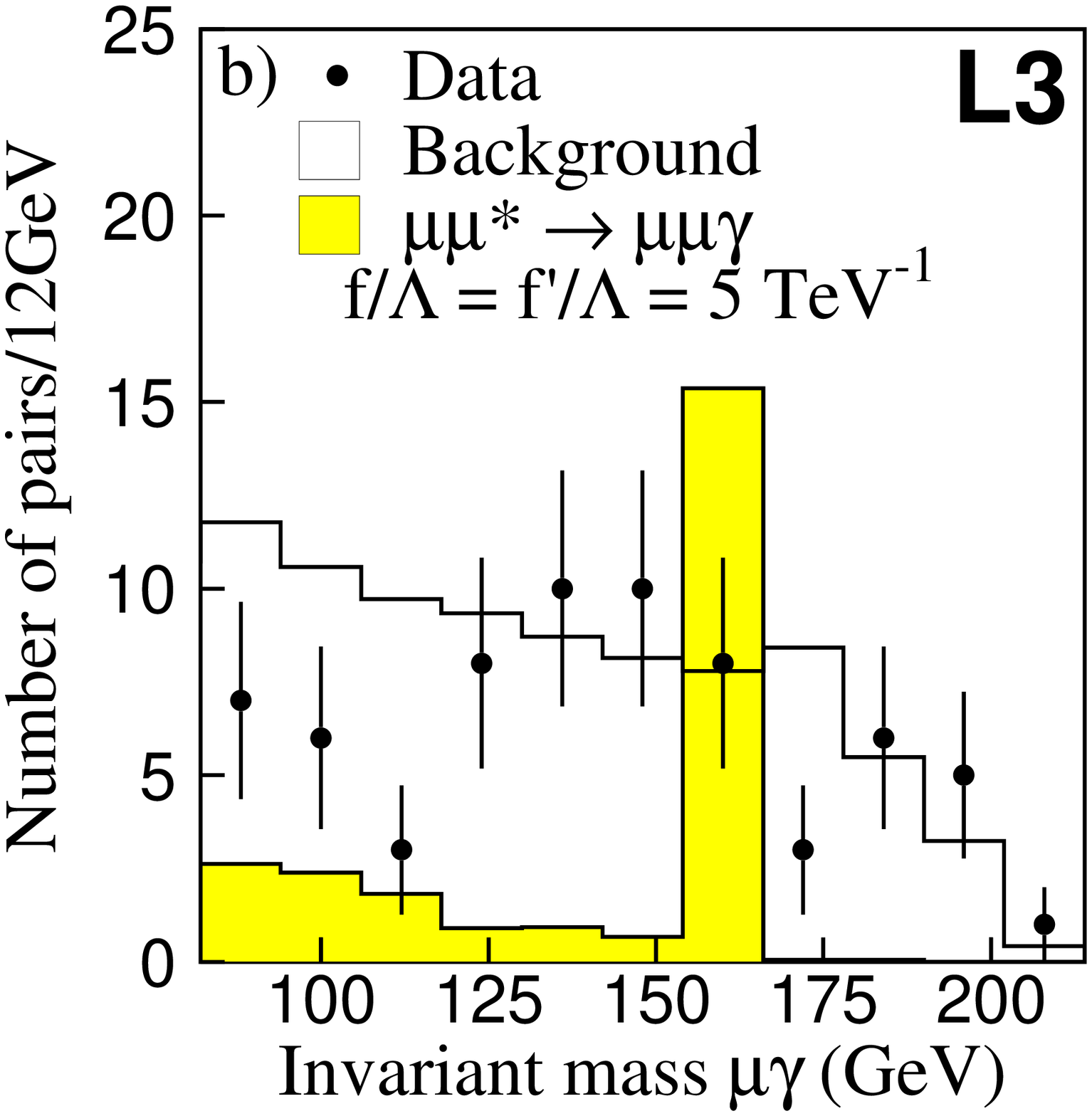}
    \includegraphics[width=0.49\textwidth]{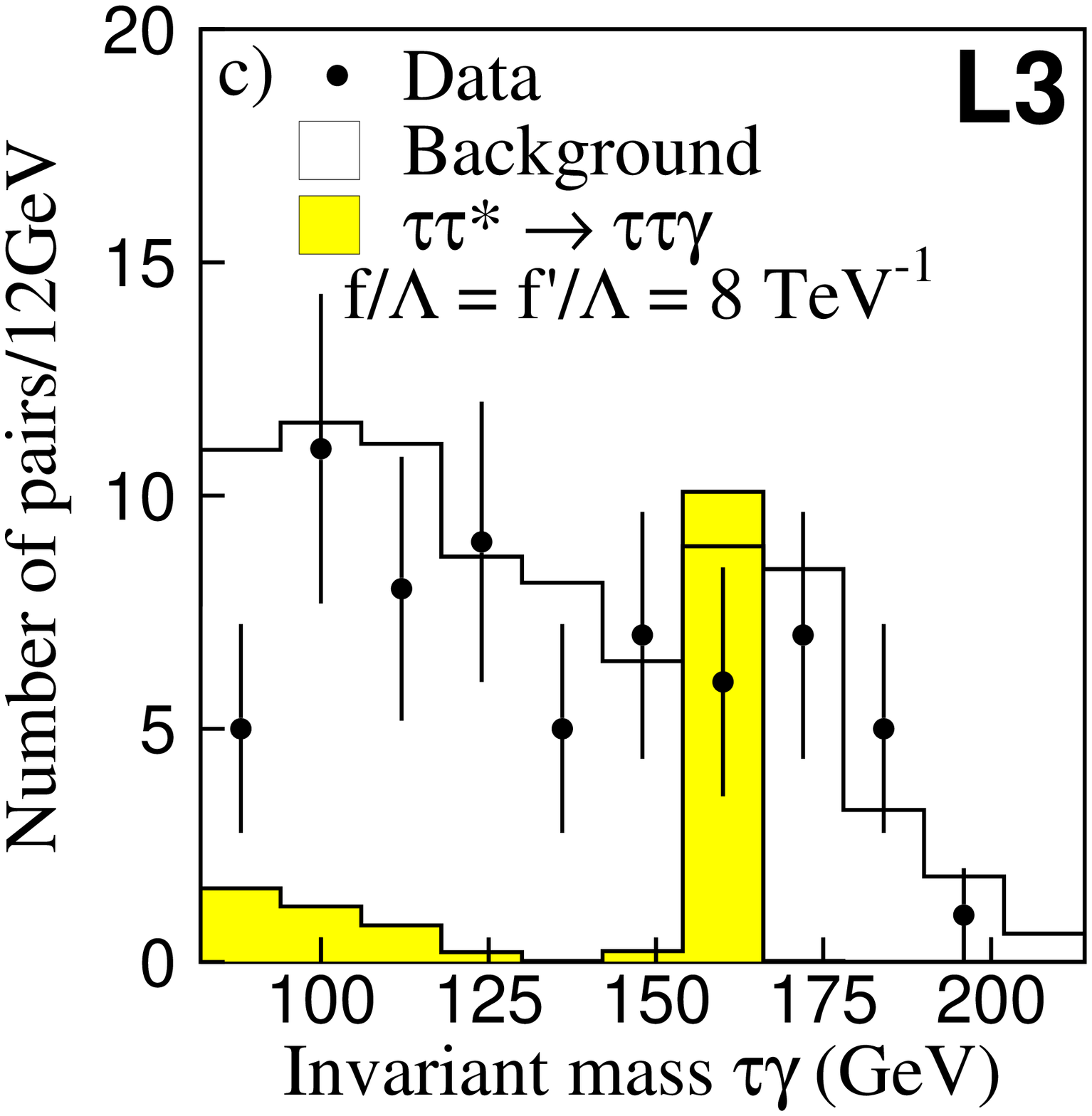}
    \includegraphics[width=0.49\textwidth]{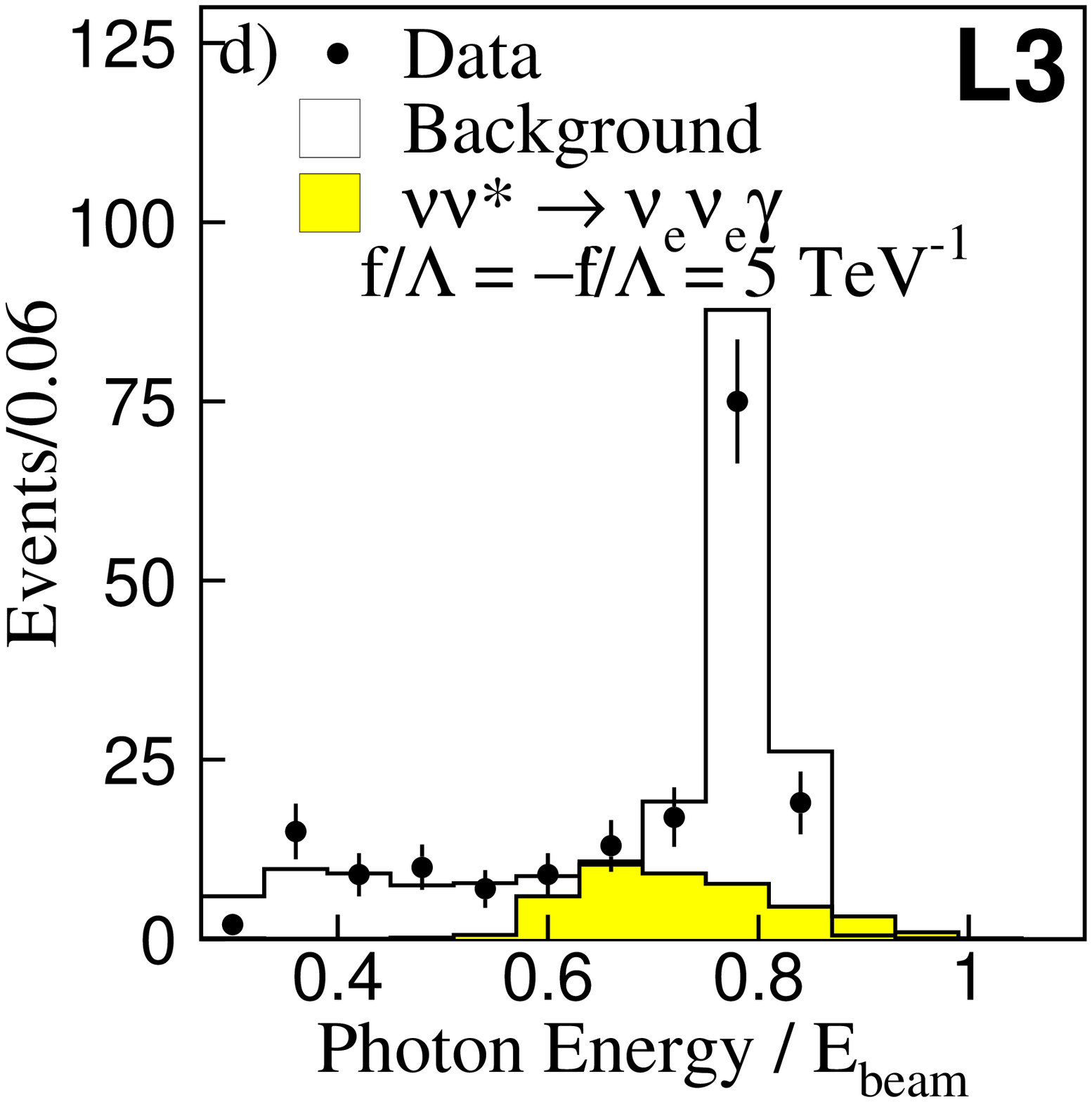}
  \end{center}
  \icaption{The invariant mass distributions for the a) e$-\gamma$,
           b) $\mu-\gamma$, and c) $\tau-\gamma$ systems. 
           d) The normalised energy distribution
           of single photon events.
           The expected signal
           for a singly-produced excited lepton with a mass of $160 \GeV$  
           is shown together with data and the Standard
           Model background expectation, due to fermion
           pair-production with initial state radiation photons.
           The signals are plotted for the arbitrary choice of couplings 
           displayed in the Figures.
  \label{fig:rad}}
\end{figure}
\clearpage

\begin{figure}[htb]
  \begin{center}
    \includegraphics[width=0.40\textwidth]{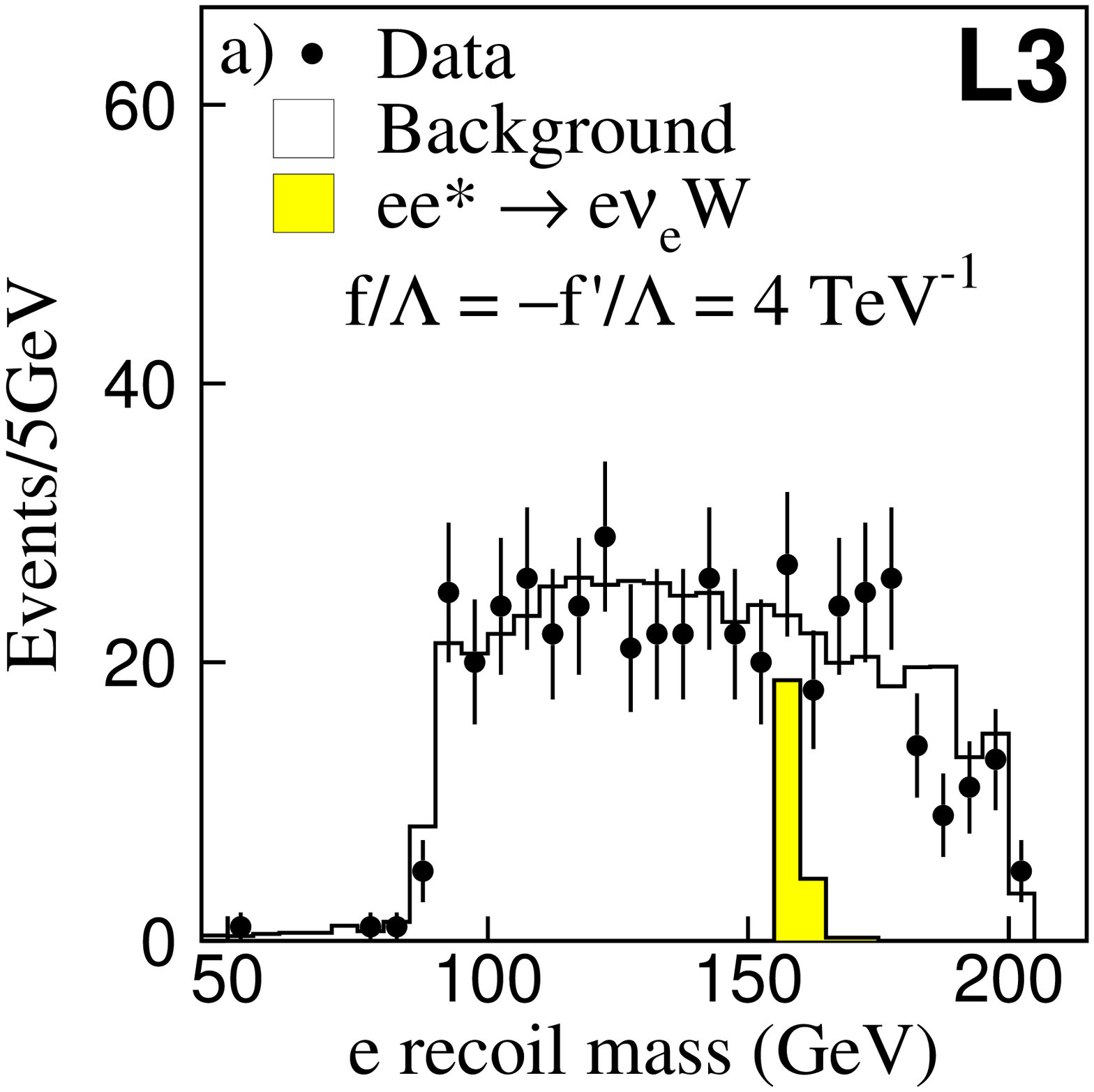}
    \includegraphics[width=0.40\textwidth]{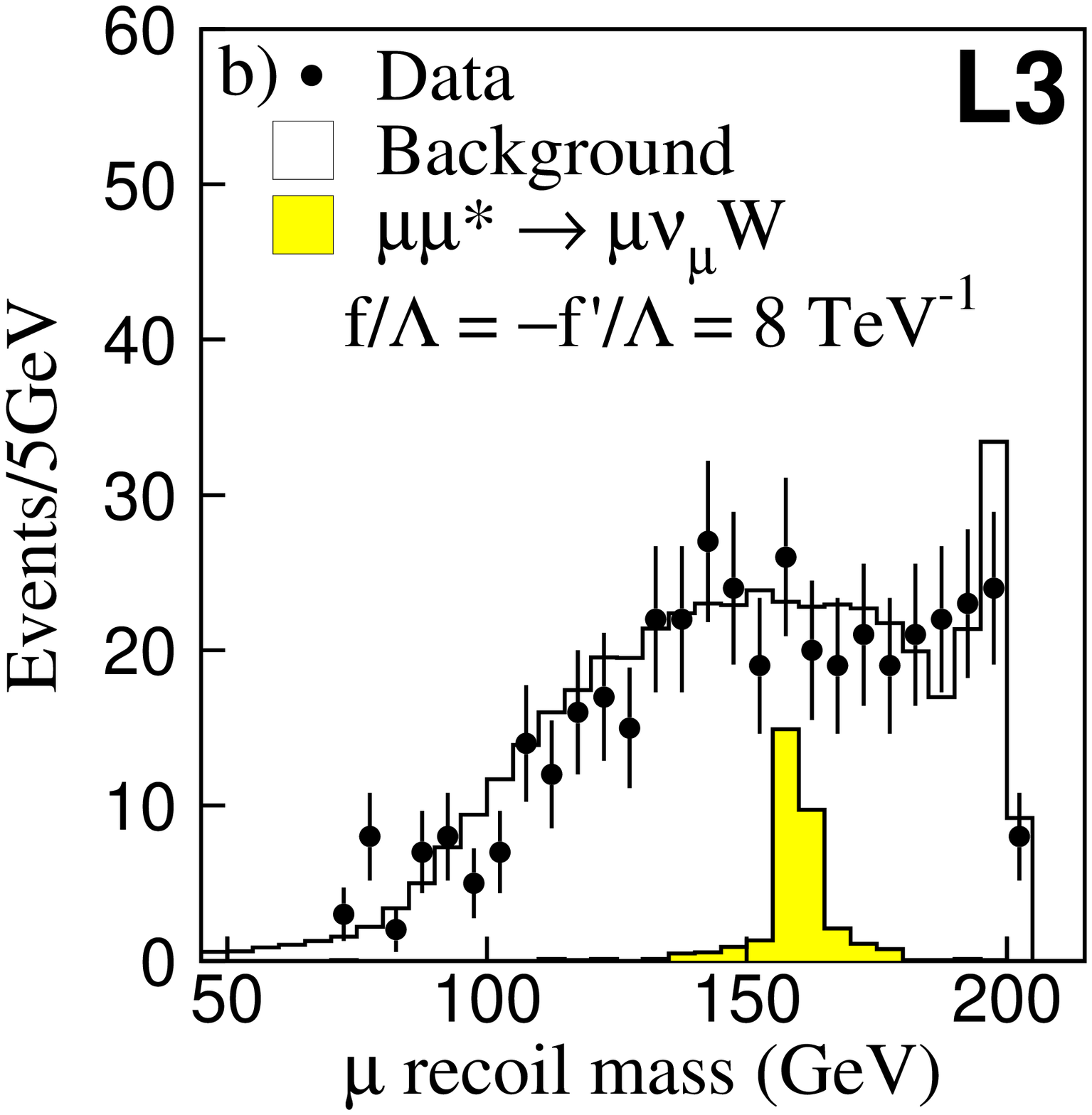}
    \includegraphics[width=0.40\textwidth]{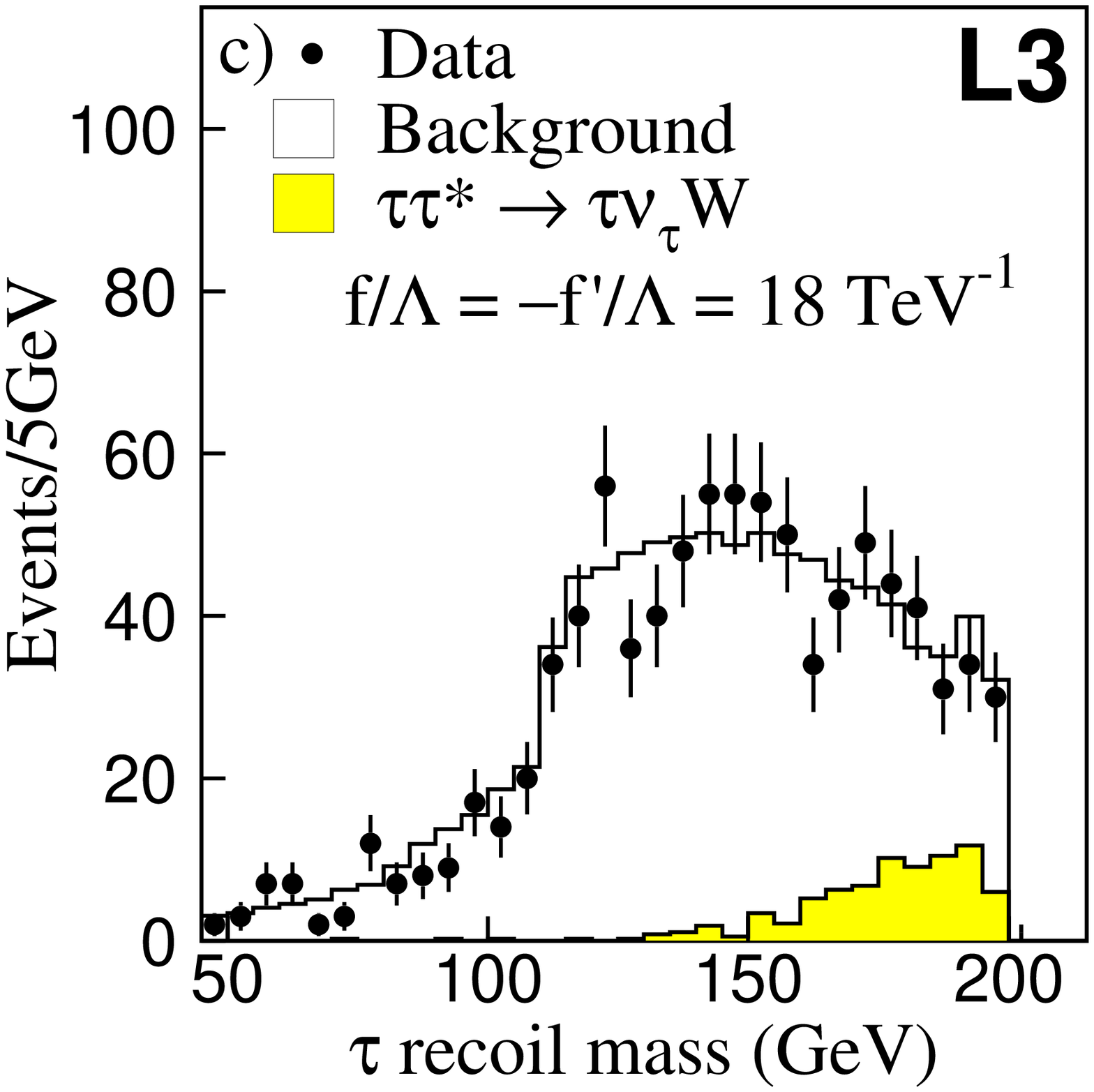}
    \includegraphics[width=0.40\textwidth]{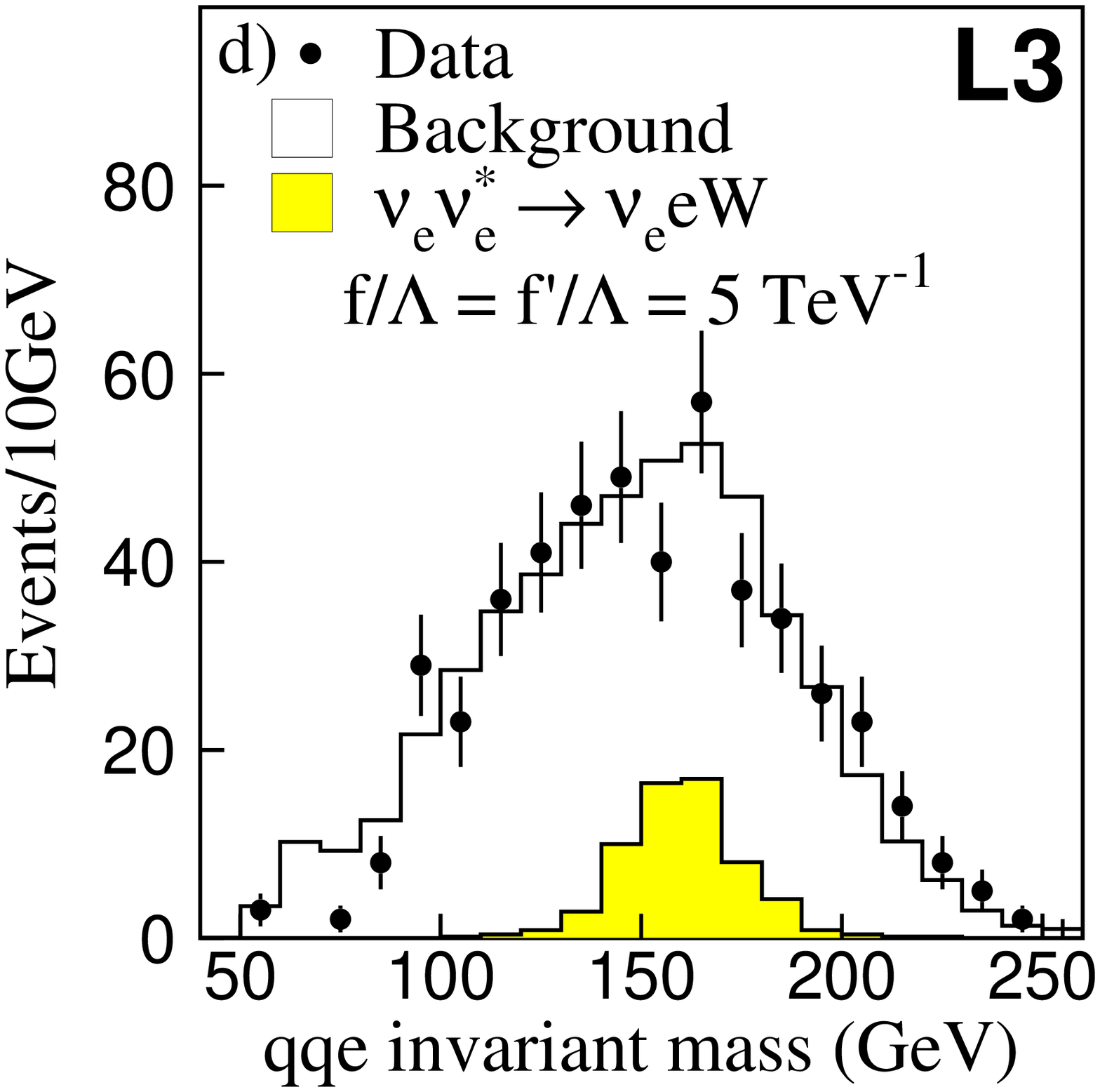}
    \includegraphics[width=0.40\textwidth]{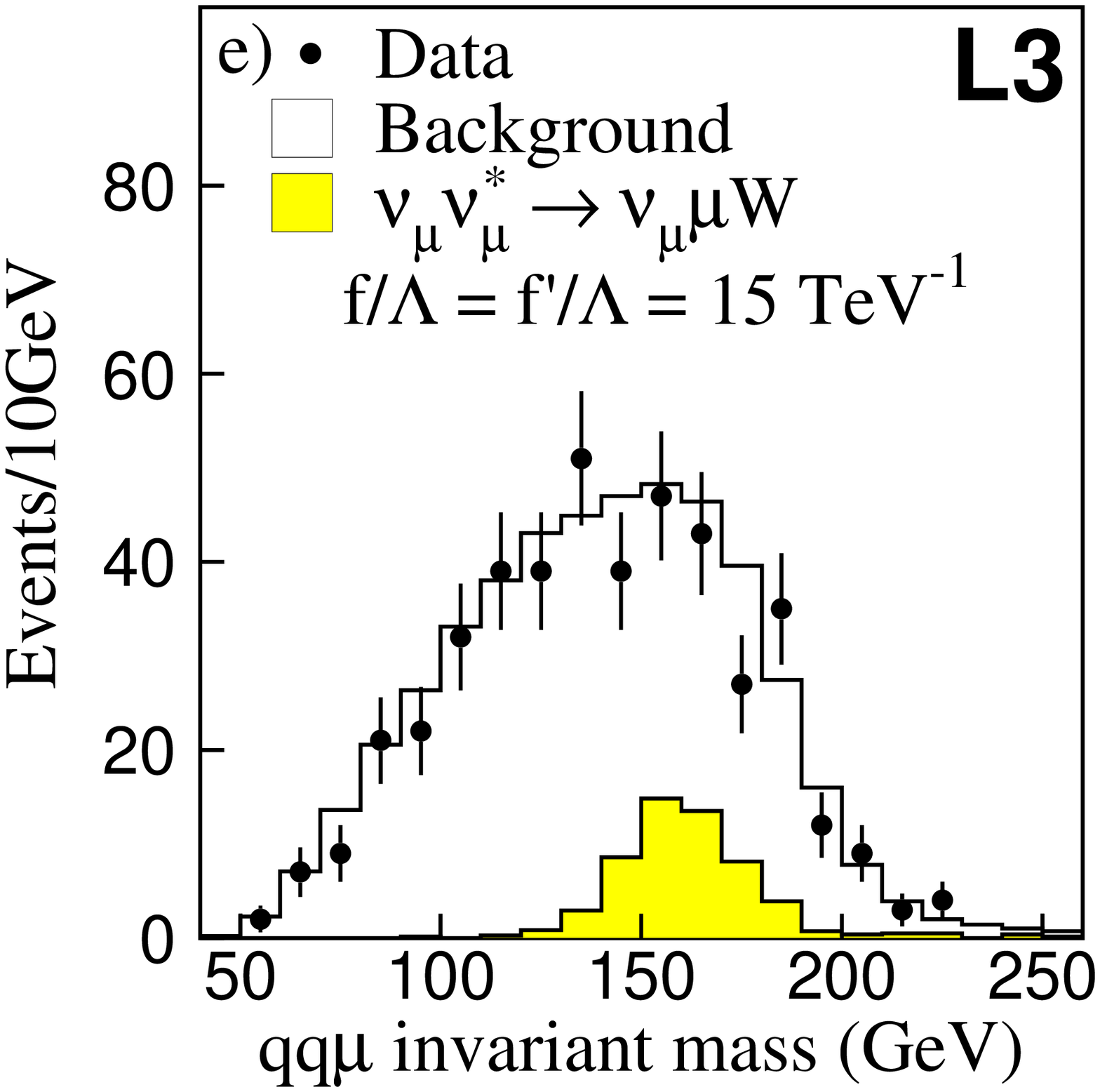}
    \includegraphics[width=0.40\textwidth]{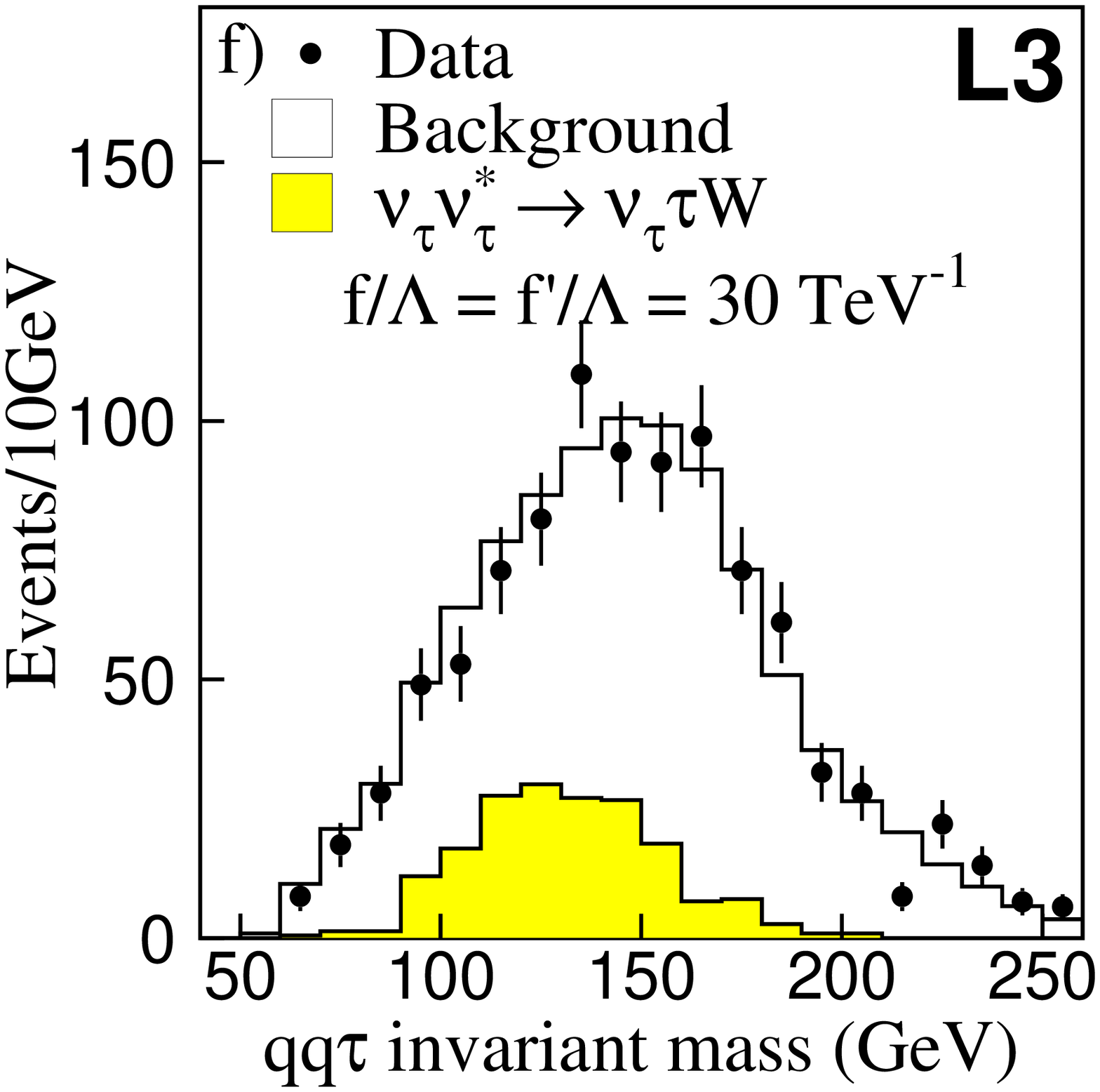}
  \end{center}
  \icaption{Lepton recoil mass distributions for the 
           a) $\rm qqe\nu$, b) $\rm qq\mu\nu$
           and c) $\rm qq\tau\nu$ selections. 
           Invariant mass distributions for the d) $\rm qqe\nu$, 
           e) $\rm qq\mu\nu$ and f) $\rm qq\tau\nu$  selections.
           The expected signal
           for a singly-produced excited lepton with a mass of $160 \GeV$ 
           is shown together with data and Standard
           Model background expectations dominated by charged-current
           four-fermion production.
           The signals are plotted for the arbitrary choice of couplings 
           displayed in the Figures.
  \label{fig:weak}}
\end{figure}
\clearpage

\begin{figure}[htb]
  \begin{center}
    \includegraphics[width=0.49\textwidth]{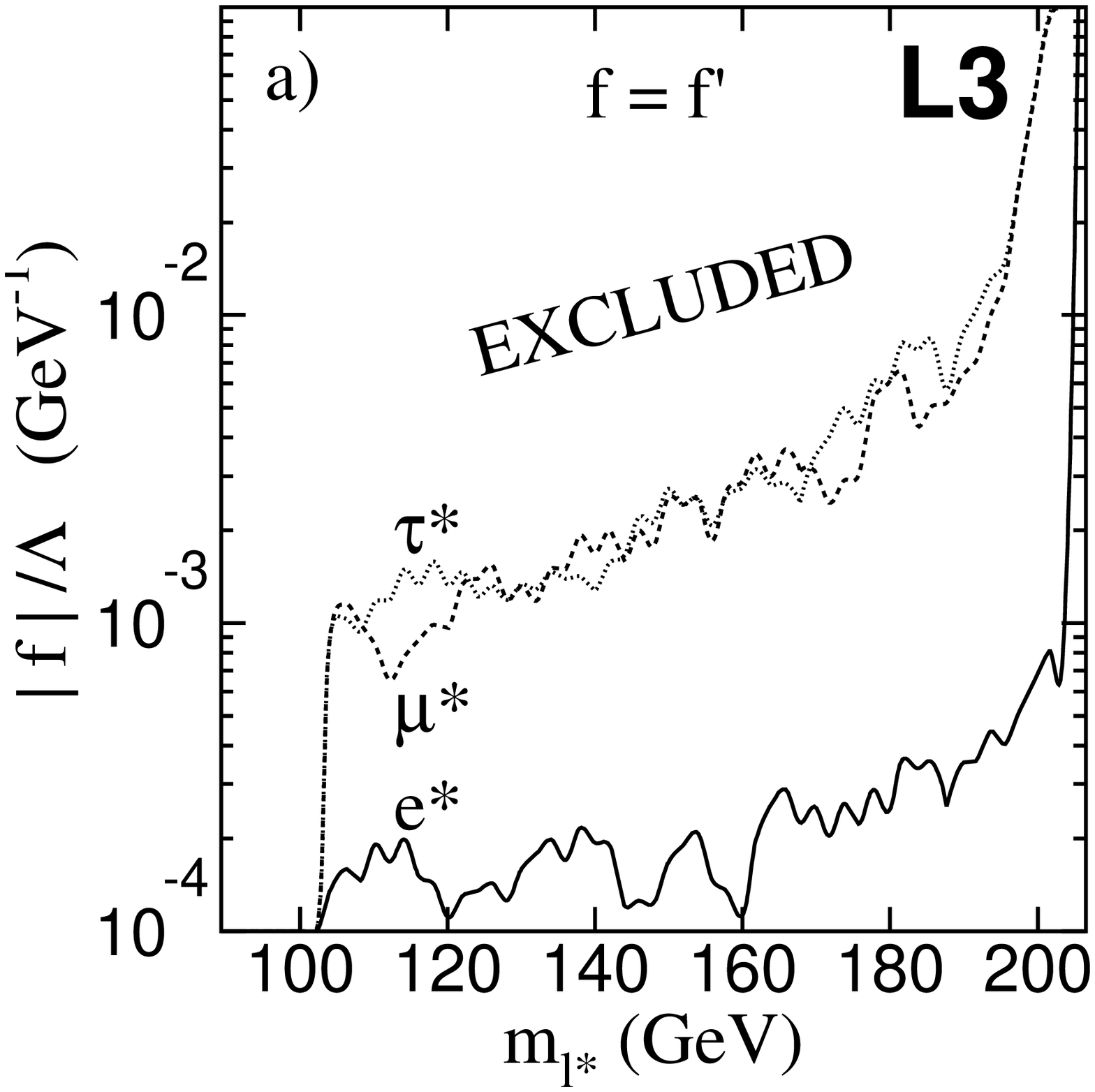}
    \includegraphics[width=0.49\textwidth]{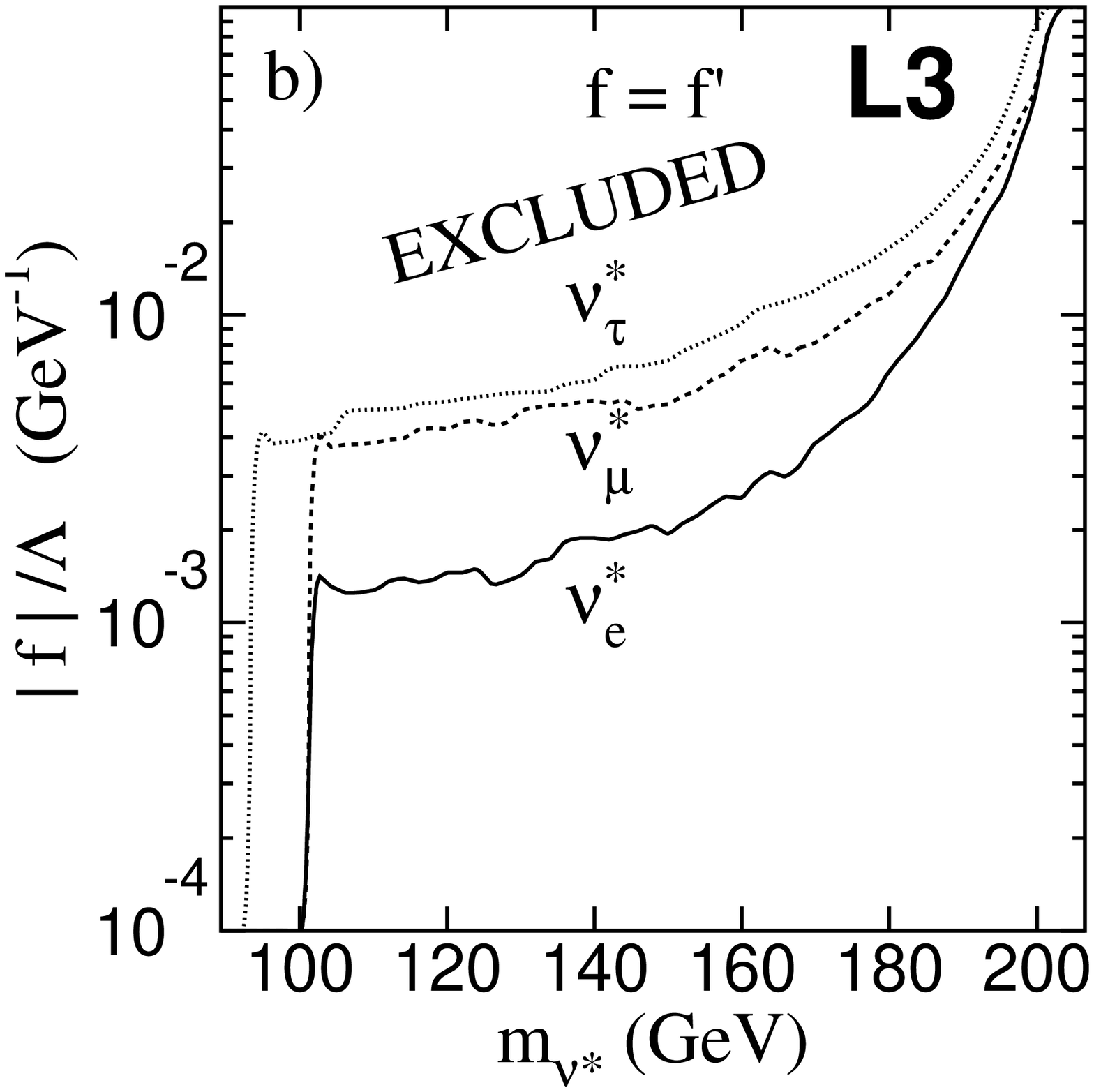}
    \includegraphics[width=0.49\textwidth]{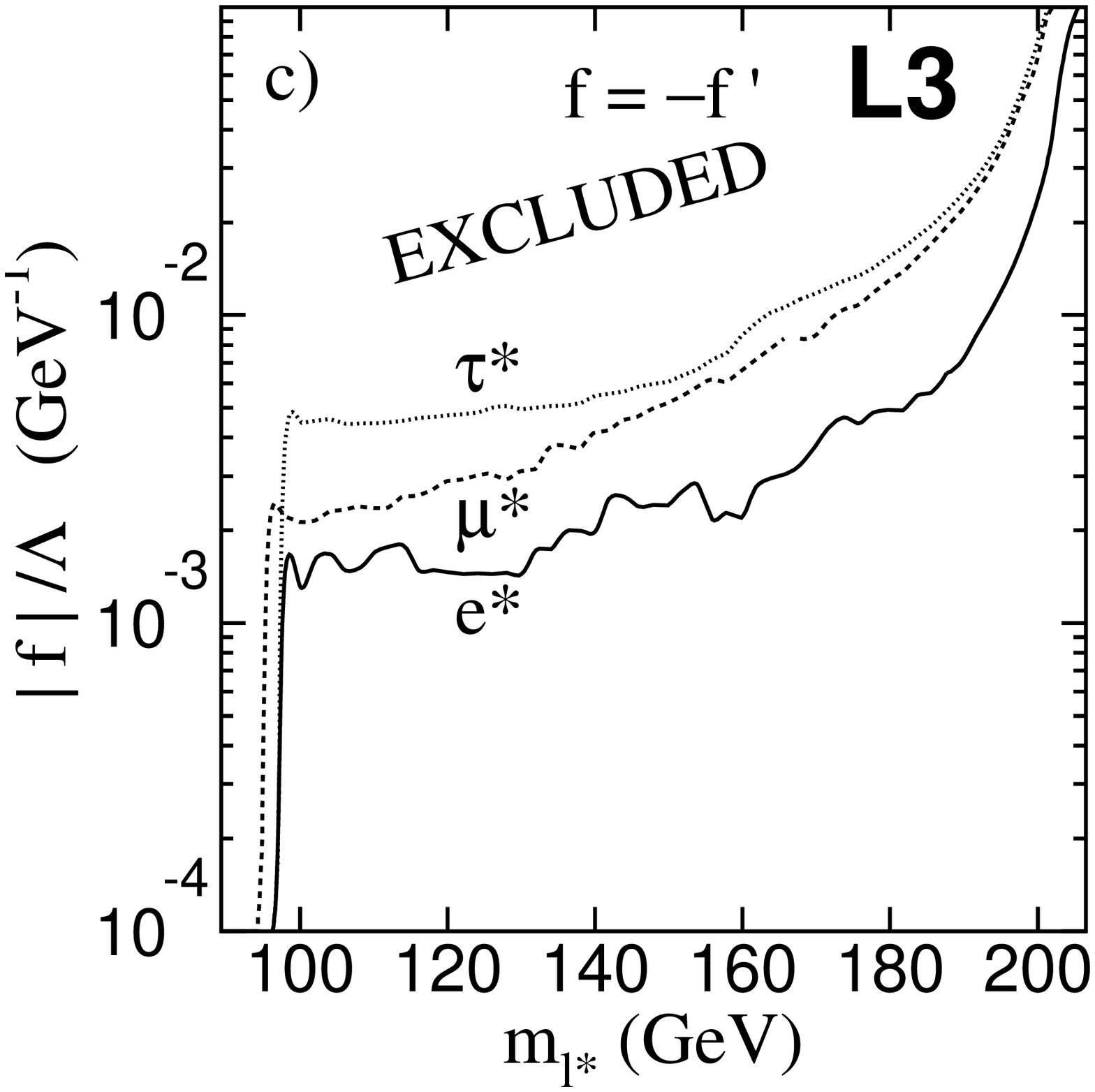}
    \includegraphics[width=0.49\textwidth]{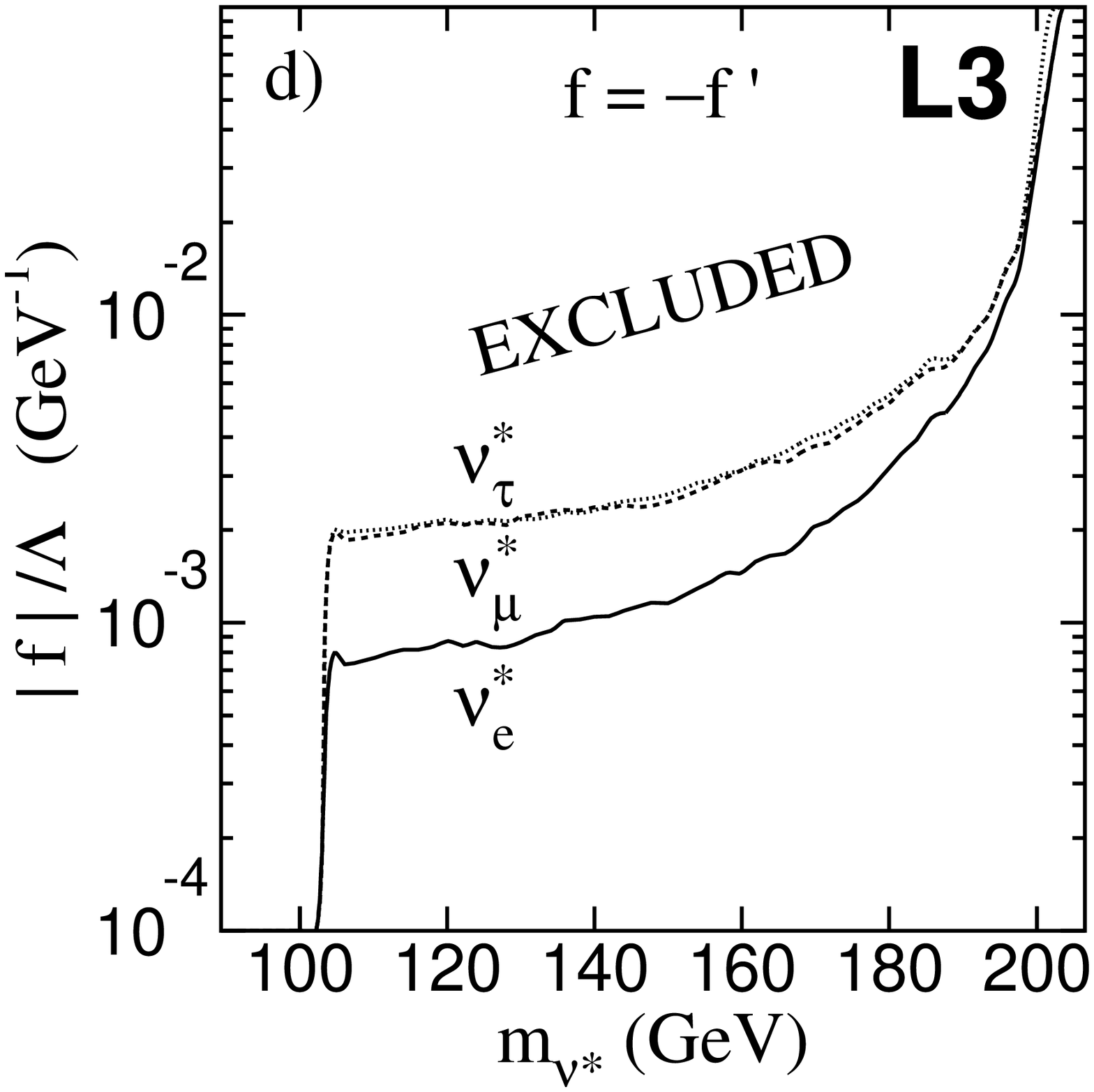}
  \end{center}
  \icaption{95\% confidence level upper limits on  
           ${\textstyle |f|} / {\textstyle \Lambda}$,
           as a function of the excited lepton mass  with $f=f'$ for
       a) $\rm e^*$, $\mu^*$ and $\tau^*$,
       b) $\rm \nu_e^*$, $\nu_{\mu}^*$ and $\nu_{\tau}^*$,
           and with $f=-f'$ for
       c) $\rm e^*$, $\mu^*$ and $\tau^*$,
       d) $\rm \nu_e^*$, $\nu_{\mu}^*$ and $\nu_{\tau}^*$.
  \label{fig:limites}}
\end{figure}
\clearpage

\end{document}